\pdfoutput=1

\documentclass[preprint]{aastex631}

\usepackage[T1]{fontenc}
\usepackage{amsmath}

\begin{document}

\title{BS~Cassiopeiae: A Contact Binary with Starspot-Driven Variability and Evidence for Additional Components}

\author[0000-0002-8394-7237]{Min-Ji Jeong}
\affiliation{Korea Astronomy and Space Science Institute, Daejeon 34055, Republic of Korea}
\email{mjjeong@kasi.re.kr}

\author[0000-0001-8591-4562]{Chun-Hwey Kim}
\affiliation{Department of Astronomy and Space Science, Chungbuk National University, Cheongju 28644, Republic of Korea}

\author[0000-0002-8692-2588]{Kyeongsoo Hong}
\affiliation{Korea Astronomy and Space Science Institute, Daejeon 34055, Republic of Korea}

\author[0000-0003-2772-7528]{Mi-Hwa Song}
\affiliation{Department of Astronomy and Space Science, Chungbuk National University, Cheongju 28644, Republic of Korea}

\author[0000-0002-6687-6318]{Hye-Young Kim}
\affiliation{Korea Astronomy and Space Science Institute, Daejeon 34055, Republic of Korea}

\author[0000-0001-9339-4456]{Jang-Ho Park}
\affiliation{Korea Astronomy and Space Science Institute, Daejeon 34055, Republic of Korea}

\author[0000-0002-2641-9964]{Cheongho Han}
\affiliation{Department of Physics, Chungbuk National University, Cheongju 28644, Republic of Korea}

\author{Joh-Na Yoon}
\affiliation{Department of Astronomy and Space Science, Chungbuk National University, Cheongju 28644, Republic of Korea}
\affiliation{Chungbuk National University Observatory, Chungbuk National University, Cheongju 28644, Republic of Korea}

\begin{abstract}
We present a comprehensive photometric and spectroscopic analysis of the contact eclipsing binary BS~Cas, based on 22-yr multi-site light curves and 2-yr high-resolution spectra.
The long-term photometry shows pronounced year-to-year changes in the light-curve shape, including eclipse-depth reversals between the primary and secondary minima and variations in the O'Connell effect.
The first double-lined radial-velocity (RV) curves of BS~Cas are derived from the spectra using broadening-function (BF) profiles.
Sharp peaks absent from the 2018 BFs appear in the 2020 BFs, suggesting an additional component along the same line of sight as BS~Cas.
Simultaneous modeling of the $BVR$ light curves and the RVs shows that BS Cas is a contact binary in which the more massive component is cooler than its less massive companion, with a mass ratio of 2.515 and a fill-out factor of 32.3\%.
Multi-epoch modeling suggests that the long-term photometric variations arise from evolving starspots and changes in the third-light contribution.
The eclipse-timing variation diagram confirms a secular orbital period decrease at a rate of $-2.305 \times 10^{-7}~\mathrm{days\,yr^{-1}}$ consistent with angular momentum loss and mass transfer, along with an 18.376-yr modulation interpreted as a light-travel-time effect from a third body.
The TESS timings reveal anti-correlated variations between the primary and secondary minima that appear to result from differential rotation of high-latitude starspots.
These results suggest that BS~Cas is a physically complex contact binary in which long-term photometric and eclipse-timing variations reflect multiple processes, including magnetic activity and the influence of a third body.

\end{abstract}

\keywords{
    \href{https://astrothesaurus.org/uat/208}{CCD photometry (208)}, 
    \href{https://astrothesaurus.org/uat/297}{Contact binary stars (297)}, 
    \href{https://astrothesaurus.org/uat/443}{Eclipsing binary minima timing method (443)}, 
    \href{https://astrothesaurus.org/uat/444}{Eclipsing binary stars (444)}, 
    \href{https://astrothesaurus.org/uat/1081}{Multiple stars (1081)},
    \href{https://astrothesaurus.org/uat/1332}{Radial velocity (1332)},     
    \href{https://astrothesaurus.org/uat/1558}{Spectroscopy (1558)}, 
    \href{https://astrothesaurus.org/uat/1572}{Starspots (1572)},
    \href{https://astrothesaurus.org/uat/1580}{Stellar activity (1580)}
}

\accepted{2026 September 09}

\section{Introduction} \label{sec:intro}

A late-type contact binary consists of two stars that fill their Roche lobes and share a common envelope \citep{1968ApJ...151.1123L,1968ApJ...153..877L}.
The orbital period is generally shorter than a day because the two components are in very close proximity.
The mass ratio of the less massive star to the more massive component ranges from 0.06 to 1.0, and most systems contain late-type F--K stars \citep{2001AJ....122.1974R, 2012JASS...29..145E}. 
\citet{1970VA.....12..217B,1977VA.....21..359B,1984PASP...96..646B} classified contact binaries into A- and W-subtypes based on the relationship between the mass and temperature of the components. 
In A-subtype systems, the more massive component is hotter, whereas the opposite is true for W-subtype systems. 
When the hotter star is occulted, the primary eclipse (at phase 0) becomes deeper than the secondary eclipse (at phase 0.5). 
However, because light variations caused by surface-brightness inhomogeneities, such as starspots or mass-transfer processes, can alter the eclipse depths, the relative eclipse depths provide only an initial indication of the A/W subtype.
Radial velocity curves provide a reliable mass ratio and thereby identify the more massive component, a constraint that is particularly valuable for systems that do not exhibit total eclipses.
Combined with the relative temperatures of the components derived from light-curve modeling, the spectroscopic mass ratio provides a more robust basis for the physical interpretation of the A/W subtype.
Based on a statistical investigation of 159 contact binaries whose absolute physical quantities are well determined from photometric and spectroscopic observations, \citet{2024AJ....167..280J} reported that the current sample size is still insufficient to determine which of the currently proposed evolutionary scenarios best explains the evolutionary link between A-subtype and W-subtype contact binaries.
They concluded that the sample size should be enlarged by including systems with accurately determined spectroscopic mass ratios and high-quality light curves to better constrain the evolutionary relationship between A- and W-subtype contact binaries.

Many contact binary systems with a convective envelope exhibit magnetic activity. 
This activity is known to produce cool starspots that can introduce pronounced light-curve distortions.
The most representative phenomenon caused by these starspots is the O’Connell effect, which is characterized by the maximum light level following the primary eclipse being different from that following the secondary eclipse.  
The starspots can also cause eclipse-depth reversal, in which the secondary eclipse becomes deeper than the primary. 
Such reversals have been reported in TY~UMa \citep{2002MNRAS.331..707K}, GN~Boo \citep{2015AJ....149..164W}, TYC~4002-2628-1 \citep{2022MNRAS.517.1928G}, and V2790~Ori \citep{2026ApJ...997..289W} based on multi-epoch light curves.
The subtypes of contact binaries exhibiting eclipse-depth reversals have been found to alternate between the A- and W-subtypes over time \citep{2024ApJ...975..231X, 2026ApJ...997..270W, 2026ApJS..282...17Z}. 
\citet{2024ApJ...975..231X} listed eight contact binaries whose subtypes are known to show such variability. 
Through light-curve analyses of V694 Peg, \citet{2024ApJ...975..231X} demonstrated that dark spots can produce acceptable photometric solutions under different subtype assumptions, suggesting that the subtype may not change physically but instead only appear to vary due to morphological effects. 
Similarly, \citet{2026ApJ...997..270W}, based on an analysis of multi-epoch light curves of CN Tri, showed that although CN Tri sometimes exhibits an A-subtype-like light-curve morphology owing to strong starspot activity, it is intrinsically a W-subtype contact binary. 
\citet{2026ApJS..282...17Z} suggested that magnetic activity is the primary driver of the observational phenomena seen in W-subtype systems (eclipse-depth reversals and variable O’Connell effects) and that such activity may explain the observed apparent transitions between the A- and W-subtypes. 
Therefore, \citet{2022MNRAS.517.1928G} cautioned that care must be taken when determining the subtype of contact binaries based solely on light-curve morphology. 
Likewise, \citet{2024ApJ...975..231X} emphasized the importance of obtaining double-lined radial-velocity (RV) curves to constrain the mass ratio and reduce ambiguities in the photometric subtype classification. 
Taken together, these studies motivate joint modeling of multi-epoch light curves and RVs, which can provide more robust constraints on system parameters and photometric subtype classification.

Observational studies of contact binaries strongly support theoretical models of close binary formation, suggesting that many contact binaries belong to multiple systems, such as triples or quadruples.
These models include von Zeipel--Lidov--Kozai (ZLK) cycles accompanied by tidal friction \citep{1910AN....183..345V,  1962P&SS....9..719L, 1962AJ.....67..591K} and the magnetic braking--tidal friction mechanism \citep{2001ApJ...562.1012E, 2012JASS...29..145E}, in which magnetic braking operates in combination with tidal dissipation.
If additional companions are gravitationally bound, the center of mass of the inner binary orbits the common barycenter of the multiple system. 
Through secular gravitational torques, the inner and outer orbits can exchange angular momentum, whereas the total angular momentum of the multiple system is conserved in the absence of mass loss or external torques. 
Such effects can also influence the long-term dynamical evolution and stability of the inner binary relative to those of an isolated binary or an optically associated but unbound system.
Therefore, to understand the long-term dynamical stability of such systems, it is important to assess the number, physical properties, and gravitational binding of additional companions.

Additional companions can be identified either spectroscopically, through extra sets of spectral lines or long-term variations in the systemic velocity ($V_\mathrm{sys}$), or photometrically, through extra eclipses, third-light contributions ($\ell_3$), and light-travel-time (LTT) effects in eclipse timings \citep{1952ApJ...116..211I,1959AJ.....64..149I}; see also \citet{2026AJ....171...82S}.
However, unless $V_\mathrm{sys}$ varies periodically or the eclipse-timing-variation (ETV) diagram reveals an LTT-induced sinusoid that cannot be explained by the Applegate mechanism \citep{1992ApJ...385..621A} (see Section~\ref{sec:Results:ETVs-secular}), an additional object identified by these methods cannot be confirmed as gravitationally bound to the binary.
A comprehensive approach that combines results from multiple observational techniques is therefore required to identify and characterize additional companions robustly.
 
BS~Cas (GSC 03682-01152, TYC 3682-1152-1; $\alpha_{\mathrm{J2000.0}}=01^{\mathrm h}\,21^{\mathrm m}\,38\fs57$, 
$\delta_{\mathrm{J2000.0}}=+59\degr\,10\arcmin\,26\farcs97$; $B=12.34\pm0.15$, $V=11.84\pm0.13$) was discovered as a variable star by \citet{1931AN....243..115B}. 
\citet{1944MVS....82} described the star as a W UMa-type eclipsing binary exhibiting a light variation of 12.25--12.6 mag, and \citet{1947VeSon...1...43A} reported the first times of minima measured from photographic plates. 
In the 1990s, \citet{1992BAVSM..60....1H} measured times of minima using photographic observations. 
Subsequently, various researchers reported increasingly accurate times of minima using CCD cameras.
Recently, \citeauthor{2008AJ....136..594Y} (\citeyear{2008AJ....136..594Y}; Y08) and \citeauthor{2010RAA....10..569H} (\citeyear{2010RAA....10..569H}; H10) conducted independent analyses of their own light curves. 
Their analyses showed that BS~Cas is a contact binary belonging to the W-subtype of the W UMa type system. 
However, a discrepancy exists between the $R$-band light curves obtained by these researchers in the same observing season: the curves obtained by Y08 indicate a large difference of 0.10 mag between the primary and secondary minima, whereas those reported by H10 exhibit a small difference of only 0.01 mag.
In addition, Y08 reported an O’Connell effect characterized by a brightness difference between the two maxima ($\mathrm{Max~I - Max~II}$) of $-0.04$ mag, whereas H10 found no such effect.
These differences indicate that the two light curves have distinctly different shapes.
Moreover, as a result of these discrepancies, the temperature difference between the cooler and hotter components estimated by Y08 is 463 K, whereas that obtained by H10 is 190 K (approximately 2.4 times lower).
Y08 derived a mass ratio of $q = M_2/M_1 = 3.540$ from photometric light-curve modeling alone, whereas H10 reported $q = 2.719$ (hereafter, we define Star~1 and Star~2 as the less and more massive components, respectively).
H10 suggested that the differences in these light curves were likely caused by strong stellar activity, such as a hot or cool starspot.
Furthermore, from the orbital period analysis of BS~Cas, they found that its orbital period had been continuously decreasing, and Y08 additionally reported a cyclic variation with a period of approximately 13~yr.

Despite these efforts, no reliable physical parameters of BS~Cas have yet been determined, primarily due to the scarcity of spectroscopic measurements.
A more comprehensive study of the orbital period variation and its cause, incorporating additional times of minima, is required to supplement the previous investigations by Y08 and H10.

In this study, we present a multi-year photometric and spectroscopic analysis of BS~Cas. 
The paper is organized as follows.
In Section~\ref{sec:Observation}, we describe the observational data and reduction procedures.
Section~\ref{sec:PeriodINV} models the eclipse-timing variations with secular plus LTT terms.
Section~\ref{sec:Modeling} presents our simultaneous analysis of the spectra and historical light curves (2003--2024), from which we derive self-consistent physical parameters.
Section~\ref{sec:Discussion} presents our analysis of (i) the causes of the orbital-period variations, (ii) starspot activity on the contact-binary surface in the context of latitude-dependent differential rotation, and (iii) the possible natures of the additional components in light of our combined observational constraints.
Finally, Section~\ref{sec:Summary} summarizes our main results and presents our conclusions.

\section{Spectroscopic and Photometric Data Preparation}\label{sec:Observation}

\subsection{Photometric Observations and Archival Data}\label{sec:Observation:photometry}

Ground-based photometric observations for BS~Cas were performed at Chungbuk National University Observatory Jincheon Station (CbNUOJ; \citealt{2014ApJ...788..134K}), the Sobaeksan Optical Astronomical Observatory (SOAO), and the Lemmonsan Optical Astronomy Observatory (LOAO).
First, $BV$ observations of BS~Cas were obtained over 14 nights in 2011 at CbNUOJ, using a 60-cm reflector telescope equipped with an SBIG (ST-8) CCD camera.
The field of view (FOV) was $27^{'}\times18^{'}$. 
The SOAO observations were conducted on four nights in 2014, three nights in 2015, two nights in 2018, and six nights in 2020 using the 61-cm telescope equipped with an FLI 4K CCD camera.
The $R$ and $BVR$ filters were used in 2014--2018 and 2020, respectively; the FOV was approximately $15\farcm2 \times 15\farcm2$.
At LOAO, $VR$ observations were obtained over five nights between 21 and 25 October 2024 using a 1-m Ritchey--Chretien reflector equipped with an ARC 4k CCD camera; the FOV was $28\farcm1 \times 28\farcm1$.
All data were preprocessed with bias, dark, and flat-field corrections using the IRAF/CCDPROC package \citep{1986SPIE..627..733T, 1993ASPC...52..173T}. 
The IRAF/DIPHOT package \citep{1993PASP..105.1342S} was used to perform aperture photometry. 
From our observations, a total of 532 ($B$: 269 and $V$: 263), 1,290 ($B$: 211, $V$: 210, and $R$: 869), and 2,071 ($V$: 1,034, and $R$: 1,037) measurements were obtained at the CbNUOJ, SOAO, and LOAO sites, respectively.
The photometric data samples are listed in Table \ref{tab:LCdata}.

To construct the light curves, TYC 3682-320-1 ($\alpha_{\mathrm{J2000.0}}=01^{\mathrm h}\,21^{\mathrm m}\,31\fs97$, 
  $\delta_{\mathrm{J2000.0}}=+59\degr\,12\arcmin\,35\farcs55$; 
  $B=12.54\pm0.18$, $V=11.68\pm0.11$)
and TYC 3682-2297-1 
($\alpha_{\mathrm{J2000.0}}=01^{\mathrm h}\,21^{\mathrm m}\,27\fs39$, 
  $\delta_{\mathrm{J2000.0}}=+59\degr\,12\arcmin\,58\farcs73$; 
  $B=12.61\pm0.18$, $V=12.12\pm0.15$) were selected as the comparison ($C$) and check ($K$) stars, respectively, as their apparent magnitudes and color indices are similar to those of BS~Cas. 
The observational errors of the CbNUOJ data, estimated from the standard deviations of the $C-K$ differential magnitudes, were 0.0091 and 0.0072 mag in the $B$ and $V$ filters, respectively.
For the SOAO data, the photometric errors were 0.0031, 0.0029, and 0.0037 mag in $B$, $V$, and $R$, respectively, whereas those for the LOAO data were 0.0054 and 0.0058 mag in $V$ and $R$, respectively.
All uncertainties were estimated from the standard deviations of the differential magnitudes between the comparison and check stars.

\begin{figure*}[t]
\centering
\begin{minipage}[b]{0.32\textwidth}
    \centering
    \includegraphics[width=\linewidth]{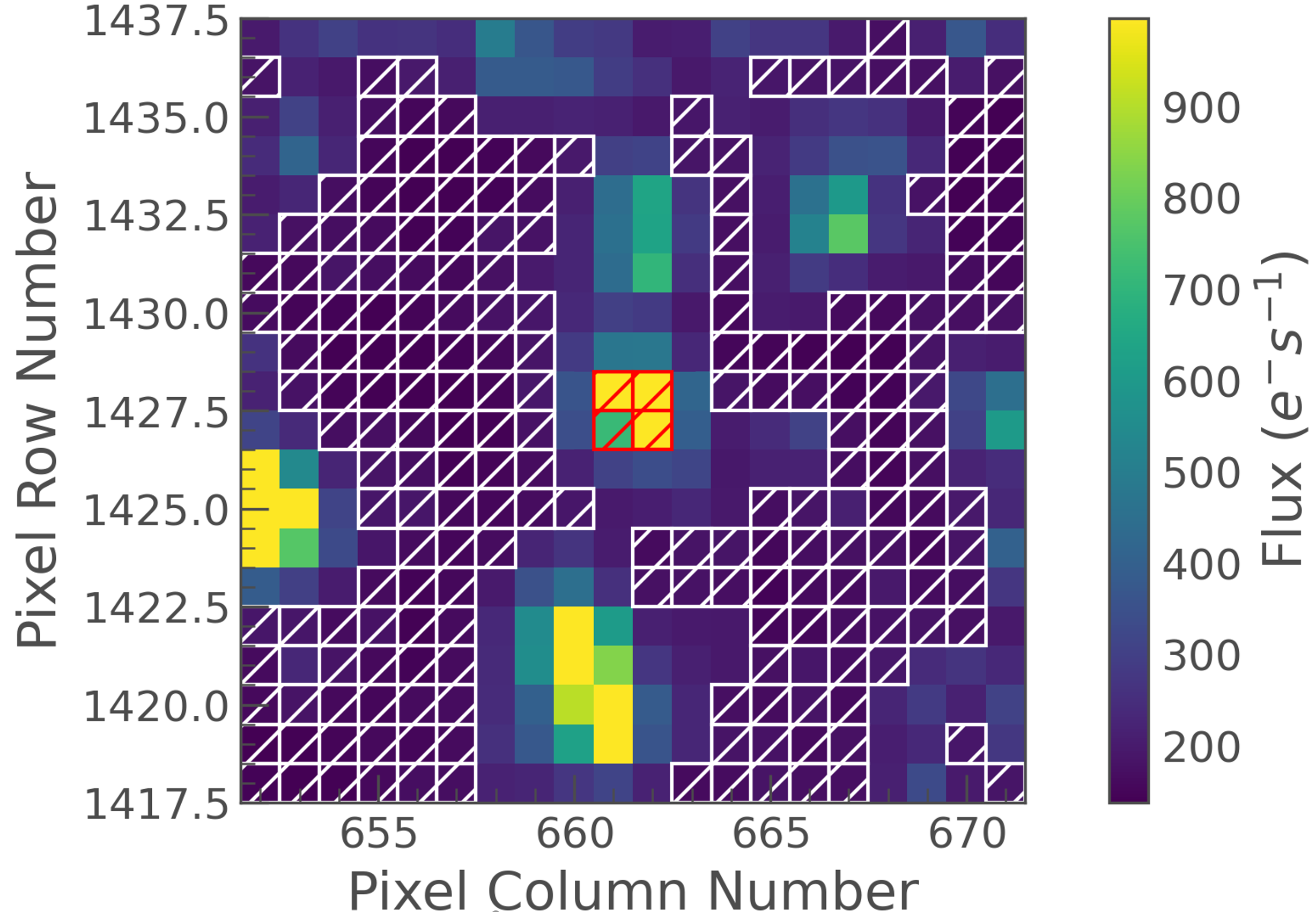}
    \par\vspace{2mm}
    (a) Sector 18
\end{minipage}
\hfill
\begin{minipage}[b]{0.32\textwidth}
    \centering
    \includegraphics[width=\linewidth]{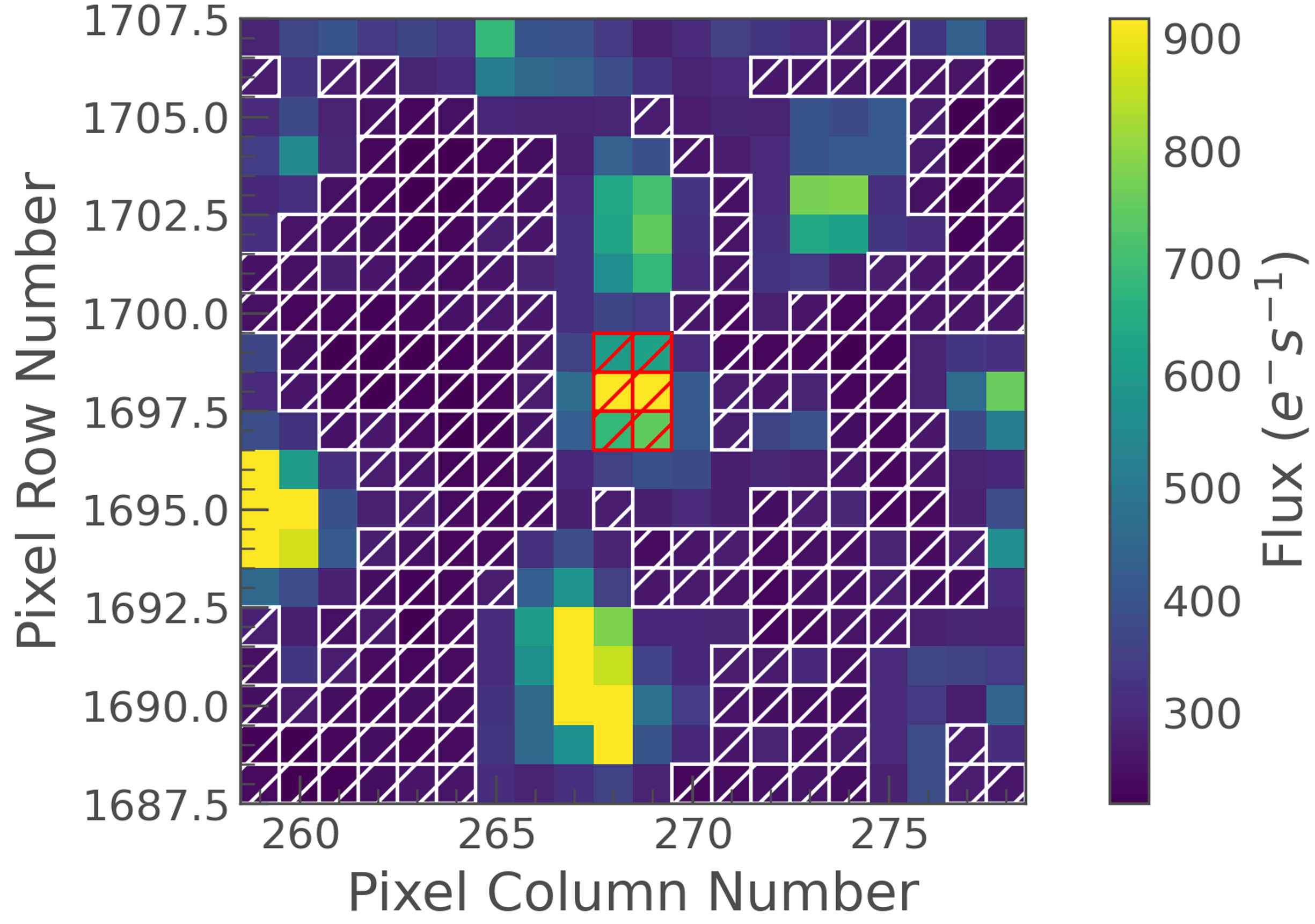}
    \par\vspace{2mm}
    (b) Sector 58
\end{minipage}
\hfill
\begin{minipage}[b]{0.32\textwidth}
    \centering
    \includegraphics[width=\linewidth]{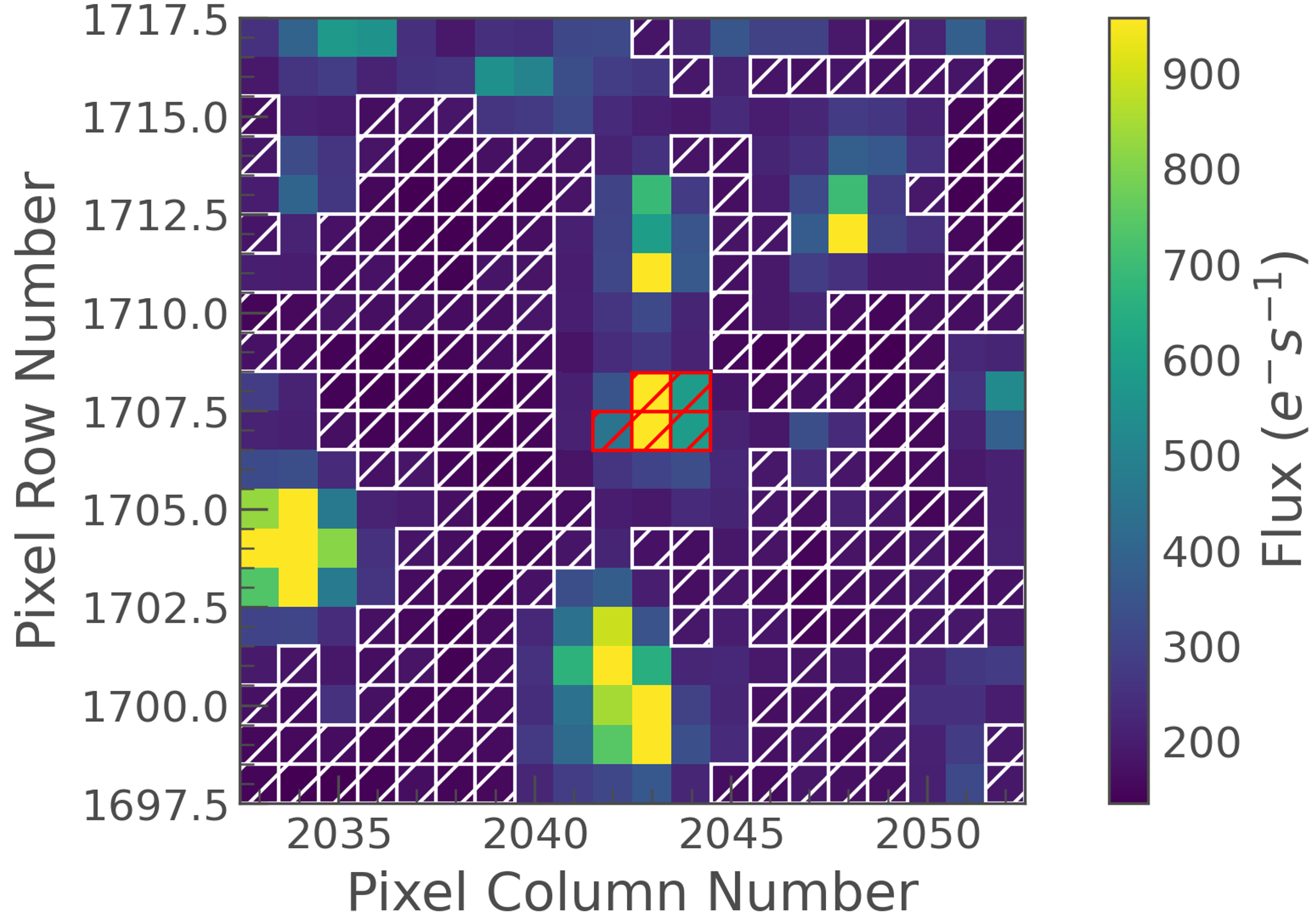}
    \par\vspace{2mm}
    (c) Sector 85
\end{minipage}
\caption{TPFs for Sectors 18, 58, and 85. The red and white rectangular boxes indicate the selected stellar mask and background pixels, respectively.}
\label{fig:TPFs}
\end{figure*}

In addition, photometric data for BS~Cas were collected from four large-scale, long-term photometric surveys.
The Transiting Exoplanet Survey Satellite \citep[TESS;][]{2015JATIS...1a4003R}, launched on April 18, 2018, has four 10.5-cm cameras that survey nearly the entire sky, which is divided into 26 sectors, and is designed to detect exoplanets.
The observation duration for each sector is approximately 27 days.
The TESS filter has a wide bandpass of 600--1000 nm, centered near the Cousins $I$ bandpass with an effective wavelength of 7865 \AA\ \citep{2015JATIS...1a4003R}.
Observations of BS~Cas were obtained in Sectors 18 (2019 November 3--27; 1425.6-s cadence), 58 (2022 October 29--November 26; 158.4-s cadence), and 85 (2024 October 26--November 21; 158.4-s cadence).
A 20-pixel cutout from the TESS full-frame image \citep[FFI;][]{2019ascl.soft05007B} centered on BS~Cas was downloaded from the Mikulski Archive for Space Telescopes (MAST) using the \texttt{search\_tesscut} function included in the \texttt{Lightkurve} package \citep{2018ascl.soft12013L}. 
Simple aperture photometry was performed using stellar masks of approximately 4--6 pixels, determined by a threshold value of 5.5. 
In this process, the background flux was evaluated as the average of approximately 200 pixels with the lowest count rate in each image frame. 
Figure \ref{fig:TPFs} presents the Target Pixel Files (TPFs), showing the stellar mask and background pixels.

The light curves were detrended and normalized following the procedure described in \citet{2024AJ....167..280J}, using the Univariate Spline fit method.
In this step, the data with high scatter and/or significant deviations from the trend were excluded.
The time values were converted to HJD$_{\rm UTC}$ using the \texttt{time}\footnote{\url{https://docs.astropy.org/en/stable/time/index.html}} module in the \texttt{astropy} package to ensure a consistent time system.
A total of 23,381 measurements were obtained. 
The time-scaled light curves are shown in Figure \ref{fig:LCtess_time}.

The Optical Monitoring Camera \citep[OMC;][]{2003A&A...411L...1W, 2012A&A...548A..79A} was designed to observe the optical emission from the primary targets of the gamma-ray instruments on board the International Gamma-Ray Astrophysics Laboratory (INTEGRAL), which was launched on October 17, 2002.
The OMC also monitored many optically variable sources within its field of view.
The observations for BS~Cas were performed between March 2003 and July 2023 using the Johnson $V$ filter.
In this analysis, we adopted 2,901 data points with \(\texttt{PROBLEMS}=0\), along with an additional 42 with \(\texttt{PROBLEMS}\ne 0\) obtained during the primary eclipse on 2007 February 8--10.
The OMC time values, given in Barycentric INTEGRAL Julian Date, were converted to HJD$_{\rm UTC}$ using the procedure applied to the TESS data.

The Super Wide Angle Search for Planets \citep[SuperWASP;][]{2003ASPC..294..405S, 2006PASP..118.1407P} conducted wide-field monitoring to detect exoplanetary transits using a broad-band filter spanning 400--700 nm.
The observations for BS~Cas were conducted between August 2007 and July 2008, overlapping with the observation periods of Y08 and H10.
In this study, we used the differential brightness data between the variable and comparison stars to minimize the scatter caused by instrumental effects and to compare the light curves.

The American Association of Variable Star Observers (AAVSO)\footnote{\url{https://www.aavso.org/research-portal}} facilitates global participation in scientific research through observations and studies of variable stars.
The AAVSO time-series photometric data for BS~Cas, obtained between July 2012 and January 2020, were retrieved for this study.
Light curves with full-phase coverage from 2012 to 2014 were obtained using the $VR$ filters, while the remaining eclipse-phase data were acquired using the $V$ filter.

One objective of the present investigation was to examine the annual variability patterns of BS~Cas and their causes.
For this purpose, we partitioned the OMC and AAVSO data into contiguous datasets separated by observational gaps exceeding 60 days.
We then selected only those datasets for which the light curves included measurements at both maxima ($\mathrm{Max~I}$, $\mathrm{Max~II}$) and both minima ($\mathrm{Min~I}$, $\mathrm{Min~II}$).
All phased light curves used for light-curve modeling in this study are presented in Figure \ref{fig:LCs}.
A summary of all photometric data used in this investigation is provided in Table~\ref{tab:Info_PhotoData}.

\subsection{Spectroscopic Observations}
The first high-resolution spectroscopic observations of BS~Cas were performed at the Bohyunsan Optical Astronomy Observatory on two occasions: 2018 November 3-4 and 2020 December 28.
The 1.8-m reflector and the Bohyunsan Observatory Echelle Spectrograph (\citealt{2007PASP..119.1052K}), which spans a spectral range of 3600--10200~\AA, were used in conjunction with a CCD camera with 2048 $\times$ 4096 pixels. 
The largest fiber (300 $\mu$m) and a 2 $\times$ 2 binning mode were employed, with exposure times of 750-s. 
All images were pre-processed with the IRAF/CCDRED package and extracted to one-dimensional spectra with the IRAF/ECHELLE package \citep{1986SPIE..627..733T, 1993ASPC...52..173T}. 
During this procedure, wavelength calibration was performed using the emission lines of a Th-Ar lamp.
In total, 36 images were acquired, with an average signal-to-noise (S/N) ratio of approximately 18.8 in the 4700--5450~\AA\ region.

\section{Orbital Period Investigation} \label{sec:PeriodINV}
To investigate the orbital-period variations, 305 times of minimum light were determined from the collected photometric data using the \citet{1956BAN....12..327K} (KW) method.
These include our own measurements: 6 from CbNUOJ, 12 from SOAO, 8 from LOAO, 19 from AAVSO, 13 from SuperWASP, 3 from OMC, and 244 from TESS.
The TESS Sector 18 data have a relatively long exposure time of 1425.6~s, resulting in only about 21 data points per orbital cycle, which increases the uncertainty when determining a single time of minimum light from one cycle.
As a test, we found that grouping five consecutive cycles reduced the mean KW timing uncertainty from about 0.0020~days for the sparsely sampled individual cycles to approximately 0.00015~days, corresponding to a reduction by a factor of about 13.
On this basis, the Sector~18 observations were grouped into sets of five consecutive cycles, and one time of minimum light was determined for each group.
In addition, 144 times of minimum light were collected from the literature, the O-C gateway\footnote{Available through the VarAstro portal, operated by the Variable Star and Exoplanet Section of the Czech Astronomical Society, at \url{https://var.astro.cz/en}.}, and the TIming DAtabase at Krakow\footnote{\url{https://www.as.up.krakow.pl/ephem/}} (TIDAK; \citealt{2004AcA....54..207K}).
Hence, a total of 449 times of minimum light ($O$) were compiled (Table \ref{tab:MINdata}).
Figure \ref{fig:ETD} shows the eclipse timing diagram (ETD) of BS~Cas, constructed using the linear ephemeris given by \citet{2001aocd.book.....K}, as follows:
\begin{equation}
    C_1 = {\rm HJD}2427984.7395 + 0.^{\rm d}44047629 \times E \label{eq1}.
\end{equation}
As shown in Figure \ref{fig:ETD}, there is a long data gap between 1945 and 1990.
Nevertheless, the orbital period of BS~Cas appears to have decreased continuously over the past 80~yr. 
This long-term variation was first reported by Y08 and later confirmed by H10. 

In addition to the secular decrease, Y08 proposed a cyclical variation with a period of 13.24~yr and a small amplitude of 0.0023 days based on all available data at that time. 
The blue solid line in the upper panel of Figure \ref{fig:ETD} corresponds to the Y08 solution (Eq. 3 in Y08), whereas the second panel shows the residuals. 
From 2007 onward, the residuals in the second panel deviate conspicuously from the expected curve; this discrepancy indicates that the Y08 solution requires revision.
Therefore, we investigated the orbital-period variation to refine the Y08 solution.

As shown in the second panel of Figure \ref{fig:ETD}, the seven H10 timings (cyan symbols) systematically lie below the general trend.
The epochs of six of these H10 timings coincide with those reported by \citet{2007OEJV...74....1B, 2008OEJV...94....1B}, and the latter timings follow the systematic trend well.
The H10 timings are approximately 0.001--0.004 days earlier than those of \citet{2007OEJV...74....1B, 2008OEJV...94....1B}; this offset is consistent with the expected difference between JD and HJD.
Therefore, we assumed that the H10 timings were reported in JD and corrected them by applying the JD-HJD offset.

To determine the orbital parameters that reflect the recent times of minimum light, we combined a quadratic ephemeris representing the parabolic change with a term describing the LTT effect due to a tertiary body, expressed as:
\begin{equation}
C_2 = T_0 + P E + A E^2 + \tau_3. \label{eq2}
\end{equation}
Following the light-time formalism of \citet{1952ApJ...116..211I, 1959AJ.....64..149I}, we write $\tau_3$ as
\begin{equation}
\tau_3 =
\frac{a_{12}\sin i_3}{c}
\left[
\frac{1-e_3^2}{1+e_3\cos\nu_{12}}
\sin(\nu_{12}+\omega_{12})
+ e_3\sin\omega_{12}
\right],
\end{equation}
where $c$ is the speed of light, $a_{12}\sin i_3$ the projected semi-major axis of the eclipsing pair around the common barycenter, $e_3$ the eccentricity of the wide orbit, $\omega_{12}$ the argument of periastron, and $\nu_{12}$ the true anomaly. 
At each observed epoch, $\nu_{12}$ was calculated from $P_3$, $T_{12}$, and $e_3$ using Kepler's equation.
Thus, the LTT term includes five adjustable parameters: $a_{12}\sin i_3$, $e_3$, $\omega_{12}$, $P_3$, and $T_{12}$.
The Levenberg--Marquardt method \citep{1992nrfa.book.....P} was applied to minimize the weighted sum of squared residuals between the observed and calculated times of minimum light.
The resulting parameters, together with several related quantities, are summarized in Table \ref{tab:Period_Inv_Results}.
The coefficient of the quadratic term ($A$) listed in the table corresponds to an annual decrease in the orbital period of $-2.305\pm0.004\times10^{-7}$ days~yr$^{-1}$, which is slightly smaller than the values reported by Y08 and H10.
In the top panel of Figure \ref{fig:ETD}, the red solid and black dashed lines represent the full model and the linear plus quadratic contributions ($A$), respectively.
The $\tau_3$ residuals obtained after removing the linear and parabolic terms are plotted in the third panel of Figure \ref{fig:ETD}.
The red dashed line shows the theoretical LTT contribution, with a period ($P_3$) of $18.376\pm0.076$~yr, an orbital eccentricity ($e_3$) of $0.500\pm0.032$, and a semi-amplitude ($K$) of $0.00532\pm0.00017$ days.
The values of $P_3$ and $K$ are approximately 5.14~yr longer and 0.0030 days larger, respectively, than those reported by Y08. 
The bottom panel of the figure shows the residuals from the full terms of Equation \ref{eq2}, demonstrating that the new solution provides a satisfactory fit to the observed times of minimum light.

\section{Investigations of Spectra and Light Curves} \label{sec:Modeling}
\subsection{Spectral Investigation} \label{sec:Modeling:spec}

\subsubsection{Radial-Velocity Determination} \label{sec:Modeling:spec:RV}
The RVs ($v_1$, $v_2$) of BS~Cas were estimated using the broadening-function (BF) method \citep{1992AJ....104.1968R, 1999ASPC..185...82R,2002AJ....124.1746R}, implemented in the \texttt{RaveSpan} software \citep{2017ApJ...842..110P}. 
The BF method determines the broadening function that transforms a sharp-line template spectrum into the observed broadened spectrum of the target system \citep{2002AJ....124.1746R}. 
The resulting BF profile provides a velocity-space representation of the stellar components, from which their RVs can be measured. 
This method is particularly useful for BS~Cas because the spectral lines of the rapidly rotating components are strongly blended. 
In this study, the RVs were determined by fitting the BF peaks with rotational profile functions.
A synthetic template spectrum with a temperature of 6000 K, metallicity of $0.0$, and surface gravity of $4.5$ was adopted from the library of local thermodynamic equilibrium spectra by \citet{2005A&A...443..735C}. 
The spectral region of 4419.8--5450.0\,\AA\ was selected. 
As this region contains numerous metallic absorption lines of photospheric origin and is relatively free from strong telluric contamination, it is well suited for RV determination.

Sample BF profiles for 2018 and 2020 are shown in Figure \ref{fig:BF}. 
Interestingly, the 2018 profiles (panels a and b) exhibit only the double peak produced by the binary components, whereas the 2020 profile (panels c and d) clearly shows an additional third peak.
We examined whether this feature could arise from a template mismatch or contamination by a nearby field source.
The same template and spectral region were used for all spectra, but the third BF peak appeared only in the 2020 spectra. 
This makes it unlikely that the feature originated from a simple template mismatch.
We also checked Gaia sources around BS~Cas and did not find a plausible high-proper-motion star or sufficiently bright nearby source that could explain the persistent BF peak in the 2020 observations.
The 300-$\mu$m BOES fiber corresponds to an on-sky aperture approximately 4.3 arcsec in diameter.
In the Gaia catalog, the closest source is located approximately 5 arcsec from BS~Cas and is approximately 9 mag fainter, whereas the next nearest sources are approximately 7.7 and 8.6 arcsec away and are approximately 8 and 5 mag fainter, respectively.
Given these considerations, light from a nearby source is unlikely to have entered the fiber aperture.
Even if such contamination had occurred, the nearby Gaia sources are too faint to have produced the third BF peak observed throughout the 2020 spectra.
Therefore, these BF profiles are consistent with the presence of an additional spectroscopic component along the line of sight toward BS~Cas that was not evident in the 2018 observations but appears in the 2020 spectra.

The RVs measured for the three components are listed in Table \ref{tab:RVt}.
The phased RVs shown in Figure \ref{fig:RVf} illustrate the orbital motion of the binary system and the nearly constant RV of the additional object.  
The RVs of the additional object ($v_3$) have a mean value of $-16.64~\mathrm{km~s^{-1}}$ and a standard deviation of $0.85~\mathrm{km~s^{-1}}$.
A detailed discussion of this additional object is presented in Section~\ref{sec:Results:ETV}.

\subsubsection{Temperature Estimation of the More Massive Component} \label{sec:Modeling:spec:Temperature}
The temperature of Star~2 (i.e., $T_2$) was estimated from color indices to be $6100~\mathrm{K}$ for Y08 ($B-V$) and $6250~\mathrm{K}$ for H10 ($J-H$ and $H-K$).
Recently, Gaia DR3 reported a temperature of $6828^{+25}_{-21}~\mathrm{K}$, which is approximately 728 and 578~K higher than the values reported by Y08 and H10, respectively.
Accordingly, we re-estimated $T_2$ using the spectrum obtained at HJD 2458427.21329 (corresponding to phase 0.034), when Star~1 was almost totally eclipsed by Star~2 and neither a significant Star~1 peak nor a third peak was present.

This analysis was performed using the Integrated Spectroscopic Framework \citep[\texttt{iSpec};][]{2014A&A...569A.111B, 2019MNRAS.486.2075B}, which incorporates the SPECTRUM code \citep{1994AJ....107..742G}, the ATLAS9 model atmospheres of \citet{2005MSAIS...8...14K} based on Kurucz models, the Gaia-ESO Survey linelist v6 including hyper-fine splitting (HFS) or isotopic shifts (ISO) (\texttt{GESv6\_atom\_hfs\_iso.420\_920nm}; \citealt{2021A&A...645A.106H})\footnote{\url{http://ges.roe.ac.uk/docs/GES_linelist_v6_report.pdf}}, and the solar abundance of \citet{2009ARA&A..47..481A}.
Based on statistical analyses, \citet{2023NatSR..1321648A} showed that the metallicity of contact binaries is distributed over a wide range from $-2.27$ to $0.57$, with the mode of $\mathrm{[Fe/H]} = 0.08$.
In a similar study, \citet{2013AJ....146...70R} reported a median metallicity of $\mathrm{[M/H]} = 0.04$.
Therefore, we adopted a metallicity of $\mathrm{[M/H]} = 0.00$ and an $\alpha$-element enhancement of $\mathrm{[\alpha/Fe]} = 0.00$.
A limb-darkening coefficient of 0.60 was adopted, assuming the solar value to be representative of a main-sequence star \citep{2011AcA....61..139S, 2024AJ....167..280J}.
The surface gravity ($\log{g_2}$) and projected rotational velocity ($v_{2}\sin{i}$) were fixed at 4.274 and 176 $\mathrm{km\,s^{-1}}$, respectively, based on a preliminary analysis of the RV and light curves discussed in Section \ref{sec:Modeling:WD}.
The projected rotational velocity was adopted under the assumption of synchronous rotation in contact binary systems.
The micro- and macroturbulent velocities ($v_{\mathrm{mic}}$, $v_{\mathrm{mac}}$) were adopted from the empirical relations of the Gaia-ESO Survey \citep{2019MNRAS.486.2075B, 2014A&A...569A.111B}.

The spectral modeling was performed over the wavelength ranges 4256.5--4501.5\,\AA\ and 4945.0--5190.0\,\AA. 
These ranges include the G-band, \ion{Fe}{1} $\lambda4271$, \ion{Fe}{1} $\lambda4383$, \ion{Mg}{2} $\lambda4481$, \ion{Fe}{1} $\lambda4957$, and the \ion{Mg}{1}b triplet. 
These features are well-established temperature indicators for dwarf stars, as documented in the Digital Spectral Classification Atlas of R. O. Gray\footnote{\url{http://www.appstate.edu/~grayro/mkclass/dsa3.pdf}} and \citet{2003ASSL..299.....L}.
As a result, $T_2$ was estimated to be $6307 \pm 73~\mathrm{K}$. 
The $v_{\mathrm{mic}}$ and $v_{\mathrm{mac}}$ were determined to be approximately 1.31 and 7.41 $\mathrm{km\,s^{-1}}$, respectively.
The best-fit model spectra from iSpec are overplotted as red solid lines in Figure~\ref{fig:Spec_Results}.
This spectroscopic estimate of $T_2$ is intermediate between the values adopted by Y08/H10 and the Gaia DR3 value, and we adopt it in the subsequent analysis.

\subsection{Light-Curve Properties and Annual Variations} \label{sec:Modeling:LCproperty}
To explore the light-curve variations of BS~Cas, we overplotted the CbNUOJ, SOAO, and LOAO light curves in Figure~\ref{fig:LCour}, all of which were obtained using the same comparison star.
The upper panel shows the phase-folded $BVR$ light curves obtained between 2011 and 2024, while the lower panel presents the differential magnitudes between the comparison ($C$) and check ($K$) stars. 
All light curves shown in the upper panel exhibit a flat bottom lasting approximately 42 minutes during the primary eclipses, whereas the secondary eclipses show rounded profiles. 
This indicates that the primary eclipses are total, whereas the secondary eclipses are annular.
Most notably, the $B$ and $V$ light curves obtained at CbNUOJ in 2011 are systematically fainter than the SOAO light curves obtained in 2020 at all phases, by approximately $0^{\rm m}.04$ to $0^{\rm m}.15$. 
The 2024 LOAO $V$ and $R$ light curves are also slightly fainter than those from 2020.
In contrast, the $(C-K)$ light curves in the lower panel remain nearly constant, indicating that the observed light variations of BS~Cas are intrinsic to the system. 
Furthermore, these variations appear to be driven by year-to-year changes in both the mean system brightness and the light-curve morphology.

The 2020 SOAO light curves show the reverse O’Connell effect, with $\mathrm{Max~II}$ brighter than $\mathrm{Max~I}$, whereas the usual O’Connell effect is present in the 2011 CbNUOJ and 2024 LOAO data.
Furthermore, the primary eclipse, during which only Star~2 is visible, is deeper than the secondary eclipse in 2011, whereas the opposite behavior is observed in the 2020 and 2024 light curves.
Remarkably, the depth of the total eclipse varies more strongly from epoch to epoch than that of the secondary eclipse.
These properties together suggest that Star~2 plays a leading role in the light-curve variability.

To examine the annual variation of the light curve in detail, the light values at the maxima ($\mathrm{Max~I}$ and $\mathrm{Max~II}$) and minima ($\mathrm{Min~I}$ and $\mathrm{Min~II}$) were measured from the selected light curves listed in Table~\ref{tab:Info_PhotoData}.
$\mathrm{Max~I-Min~II}$, $\mathrm{Max~II-Min~II}$, $\mathrm{Min~I-Min~II}$ ($\mathrm{\Delta Min}$), and $\mathrm{Max~I-Max~II}$ ($\mathrm{\Delta Max}$) are listed in Table~\ref{tab:MinMax}; the H10 and Y08 values were extracted from the literature.
The annual changes of $\mathrm{\Delta Min}$ and $\mathrm{\Delta Max}$ shown in Figure \ref{fig:DifMag} (a) and (b) exhibit significant variability.
These variations are particularly apparent in the $V$ and $R$ filters, which benefit from longer temporal coverage and a larger number of observations than the other filters.
As indicated in the figures, $\mathrm{\Delta Min}$ varies more strongly than $\mathrm{\Delta Max}$. 
Additionally, in the $V$ band, $\mathrm{Min~I}$ is fainter than $\mathrm{Min~II}$ during the 2007--2011 seasons, whereas the light curves from other epochs show the opposite behavior.
This behavior suggests eclipse-depth reversals in BS~Cas.

As shown in Figures~\ref{fig:DifMag}(a) and (b), the $R$-band light curve reported by Y08 exhibits a different morphology from that of H10, although both were obtained during the same observing season.
For a visual comparison, we vertically shifted the Y08, H10, and SuperWASP light curves from the same season to match their levels at Min~I and overplotted them in Figure~\ref{fig:Y08H10SWASP}.
This figure shows that the Y08 light curve displays a stronger O’Connell effect and exhibits a significantly shallower depth at Min~II compared with the H10 and SuperWASP light curves. 
The origin of this discrepancy could not be determined from the available data. 
Therefore, to maintain a homogeneous set of light curves for the simultaneous analysis, we conservatively excluded the Y08 light curves from the modeling in Section~\ref{sec:Modeling:WD}.

\subsection{Simultaneous Investigation of Light Curves and Radial-Velocity Curves}  \label{sec:Modeling:WD}
We simultaneously analyzed the double-lined RV curves and the 2020 SOAO $BVR$ light curves to derive the physical parameters of BS~Cas.
These photometric data were selected for their higher precision compared with other available datasets. 
The binary-star model employed in this study was the 2015 version of the Wilson--Devinney (WD) code \citep{1971ApJ...166..605W, 2003ASPC..298..323V}. 
This analysis was conducted in Mode 3, which is appropriate for a contact binary system.
The temperature of Star~2 ($T_2$) was fixed at 6307~K, based on the spectral modeling results presented in Section~\ref{sec:Modeling:spec}. 
The gravity-darkening coefficient ($g_1$ = $g_2$; \citealt{1967ZA.....65...89L}) and bolometric albedo ($A_1$ = $A_2$; \citealt{1969AcA....19..245R}) were set to 0.32 and 0.5, respectively, under the assumption that both components share a convective common envelope.
The square-root limb-darkening coefficients were adopted from the table of \citet{1993AJ....106.2096V}.  
The adjusted parameters included the reference epoch ($T_0$), orbital period ($P$), semi-major axis ($a$), systemic velocity ($\gamma_0$), orbital inclination ($i$), surface potential ($\Omega_1=\Omega_2$), mass ratio ($q=M_2/M_1$), temperature of Star~1 ($T_1$), bolometric luminosity of Star~1 ($\ell_1$), third light ($\ell_3$), and spot parameters.
As discussed in Section~\ref{sec:Modeling:LCproperty}, the light variation was largest during the totality of the primary eclipse, when only Star~2 was visible.
Based on this behavior, we introduced a cool spot on Star~2 in the WD modeling as a phenomenological representation of the asymmetric morphology, including the O'Connell effect, observed in the SOAO light curves.
As summarized in Table~\ref{tab:WDsolution}, the best-fit parameters indicate that BS~Cas is a typical contact binary with $T_1 = 6439\pm3$ K, $i=89.860\pm0.175\degr$, $q = 2.515\pm0.003$, and a fill-out factor of $f = 32.25\%$.
Figure~\ref{fig:3D_config} also shows the three-dimensional configurations of BS~Cas at four representative orbital phases based on the adopted cool-spot model. 
At phase 0.00, Star~1 lies completely behind Star~2 and is therefore fully eclipsed.

\begin{figure}
\epsscale{1.0}
\plotone{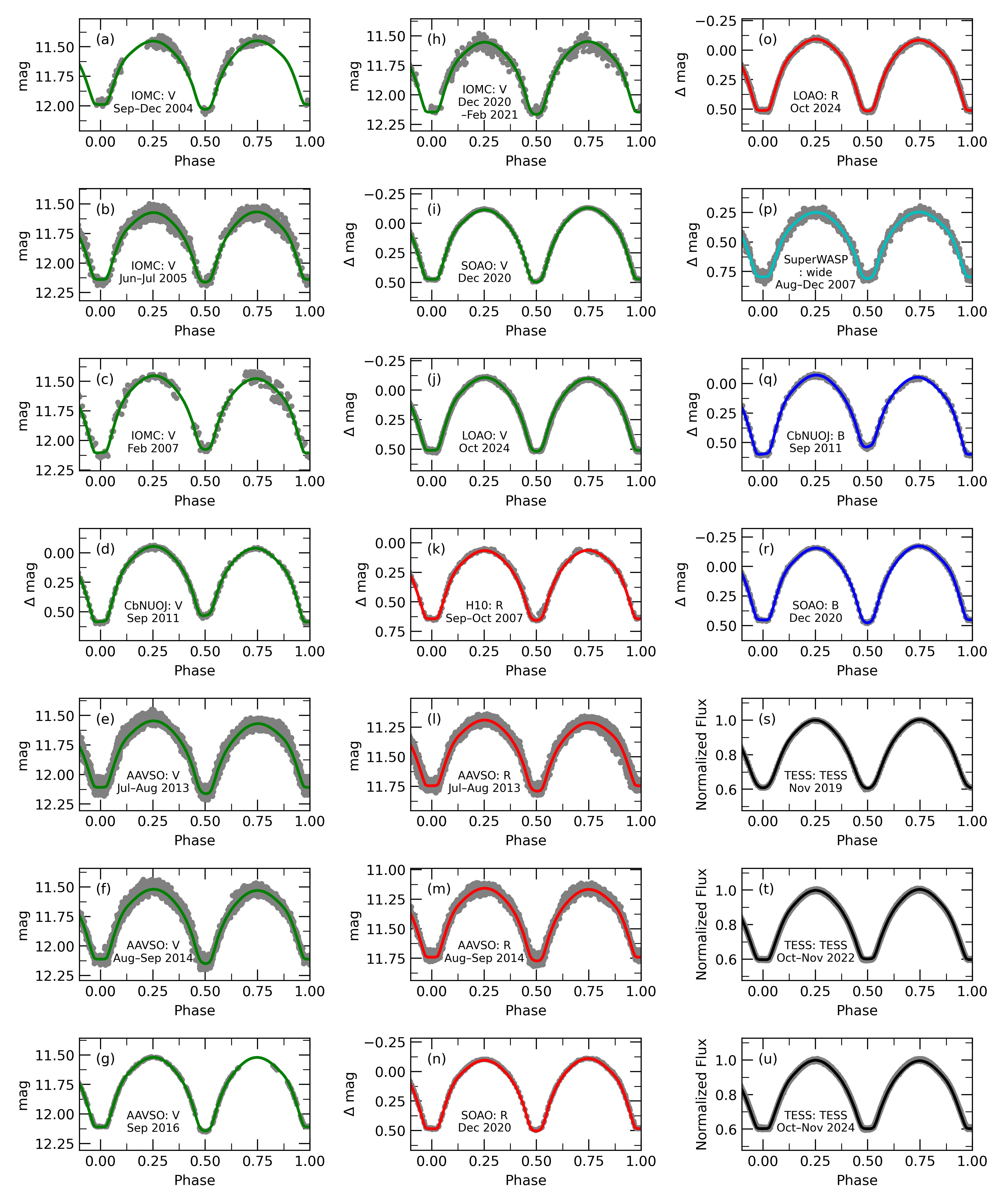}
\caption{Annual light curves of BS~Cas. 
The blue, green, red, cyan, and black solid lines represent the synthetic light curves computed with the WD spot models for the $B$, $V$, $R$, $wide$, and $TESS$ filters, respectively. 
The observed data are plotted as gray dots.}
\label{fig:LCs}
\end{figure}

\begin{figure}
\epsscale{0.7}
\plotone{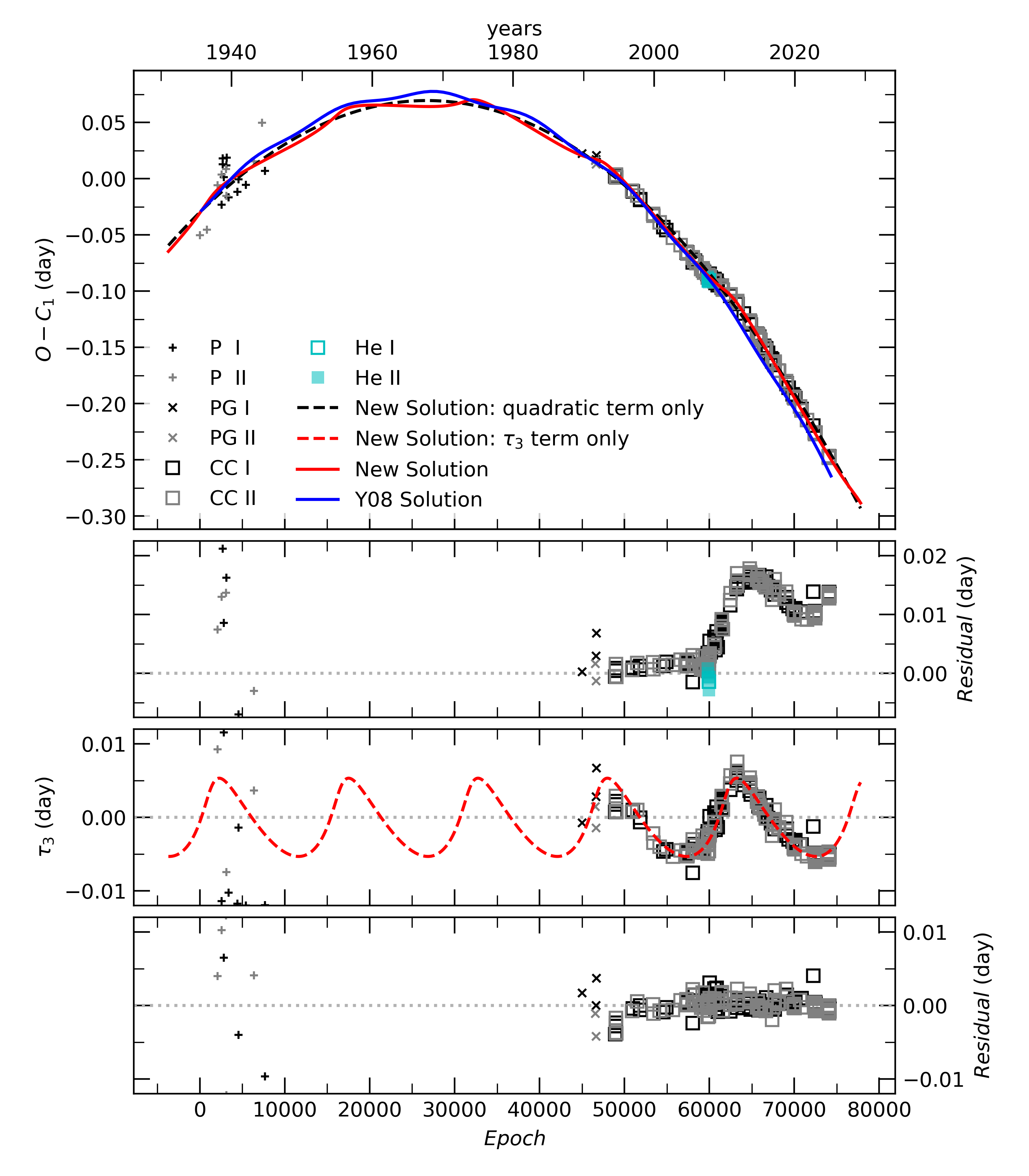}
\caption{
Eclipse-timing variation diagram of BS~Cas. 
In the top panel, the solid red curve represents the best-fitting quadratic-plus-LTT ephemeris, the dashed black curve shows the quadratic term alone, and the solid blue curve denotes the Y08 solution. 
The plus signs, crosses, and open squares represent plate, photographic, and CCD measurements, respectively. 
The black and gray symbols indicate the times of primary and secondary minima, and the timings adopted from H10 are plotted in cyan (open for primary and filled for secondary minima, respectively).
The second panel shows the differences between the observed times of minima and the Y08 solution.
The third panel shows the contribution of the LTT term, $\tau_3$, to the $O-C_1$ curve, where the red dashed line indicates the fitted $\tau_3$ modulation. 
The bottom panel displays the residuals from the final quadratic-plus-LTT ephemeris.
Some plate (P) data points lie outside the displayed y-axis ranges in the second, third, and bottom panels because of their large scatter.}
\label{fig:ETD}
\end{figure}

To investigate the long-term variations in the light curves of BS~Cas, we next applied a cool-spot model on the surface of Star~2 using the WD code and the selected photometric datasets shown in Figure \ref{fig:LCs}. 
In this procedure, only a limited set of parameters was adjusted, including $P$, $T_0$, $\ell_1$, $\ell_3$, and the spot parameters (colatitude $\theta$, longitude $\phi$, radius $r_{\rm{spot}}$, and temperature factor $T_{\rm{spot}}/T_{\rm{star}}$). 
As shown in Figure~\ref{fig:LCs}(s), the TESS light curve from Sector 18 does not clearly display a total eclipse.
This behavior is likely caused by the smearing effect associated with the long exposure time of 1425.6~s \citep{2012AJ....144...73W, 2017MNRAS.466.2488Z}; accordingly, the WD smearing factor parameter was enabled in the modeling.
By contrast, the total eclipse is clearly visible in the other TESS datasets, which have a 158.4~s exposure time.
The 2020 OMC light curve in the $V$ band was obtained during the same season as the SOAO photometric data.
Therefore, the 2020 OMC light curve was not refitted.
Instead, it was compared with the theoretical light curve computed from the WD solution presented in Table \ref{tab:WDsolution}.
All parameters were fixed except for $\ell_1$ of Star~1, which was adjusted.
As a result, the observed light curve agrees well with the theoretical light curve in overall shape.

The results are listed in Table~\ref{tab:WDsolution_All}, and the solid lines in Figure~\ref{fig:LCs} represent the corresponding model light curves.
All models except for the CbNUOJ2011 dataset have one cool spot with various radii ranges of approximately 13--40$\degr$.
The starspot parameters vary from year to year, consistent with the hypothesis that they are modulated by changes in the magnetic activity of Star~2.
In particular, the model for the CbNUOJ dataset includes two cool spots.
This implies strong starspot activity in 2011, resulting in a systematically fainter light curve at all phases than those observed in other years, as shown in Figure \ref{fig:LCour}.
In addition, the $\ell_3$ values derived from all light curves show time-dependent variations, as illustrated in Figure \ref{fig:DifMag}(c).
This variation may result from the contribution of a circumbinary object or from a gravitationally unbound star located along the same line of sight.

The absolute dimensions of BS~Cas, listed in Table~\ref{tab:WDsolution}, were calculated using the JKTABSDIM code\footnote{\url{https://www.astro.keele.ac.uk/~jkt/codes/jktabsdim.html}}\citep{2005A&A...429..645S}.
JKTABSDIM adopts the physical constants suggested by \citet{2011PASP..123..976H}.
The masses ($M$) and radii ($R$) were calculated as follows: $M_1 = 0.566\pm0.011~\rm{M_\odot}$, $M_2 = 1.423\pm0.031~\rm{M_\odot}$, $R_1 = 0.986\pm0.007~\rm{R_\odot}$, and $R_2=1.473\pm0.010~\rm{R_\odot}$.
The synchronous rotational velocities ($v_{\rm syn}$) of Star~1 and Star~2 were derived to be $v_{\rm  syn,1}=113.23\pm0.78~\rm{km~s^{-1}}$ and $v_{\rm  syn,2}=169.12\pm1.15~\rm{km~s^{-1}}$, respectively.

\section{Discussion} \label{sec:Discussion}

\subsection{Mechanisms of Secular and Periodic Changes in Orbital Period} \label{sec:Results:ETVs-secular}
BS~Cas exhibits both a secular variation and a periodic modulation in its orbital period. 
The secular decrease in the orbital period has traditionally been interpreted as a consequence of angular momentum loss \citep[AML;][]{1994ASPC...56..228B} due to magnetic braking, mass transfer from the more massive component to the less massive component, or a combination of these processes.
The orbital-period change rate expected from AML was calculated from the formula given by \citet{1994ASPC...56..228B}, using the absolute parameters presented in Table~\ref{tab:WDsolution} and adopting $k^2 = 0.1$ \citep{1976ApJ...209..829W, 2015AJ....149...93L}.
This yields $dP/dt = -7.172\times10^{-8}~\mathrm{days~yr^{-1}}$, which is approximately three times smaller than the observed value.
However, if we take into account the saturated magnetic braking discussed recently by \citet{2025ApJ...995...19F}, the orbital-period decay rate should be lower than that derived from the formula of \citet{1994ASPC...56..228B}, which excludes saturation effects.
Consequently, the value yielded by the \citet{1994ASPC...56..228B} formula can be considered an upper limit for the period decay rate. 
These considerations indicate that the AML contribution estimated from this prescription alone cannot fully account for the observed decrease in the orbital period.
The deficit may therefore be attributed to mass transfer from the more massive Star~2 to the less massive Star~1.

Assuming that the remaining orbital-period decrease, obtained by subtracting the AML-induced orbital-period change rate from the observed value, is entirely driven by fully conservative mass transfer from Star~2 to Star~1, we estimate the mass-transfer rate to be $\dot{M}_2=1.129\times10^{-7}\,\mathrm{M_\odot\,yr^{-1}}$ following the prescription of \citet{2001icbs.book.....H}. 
The absolute parameters listed in Table~\ref{tab:WDsolution} are used for this estimate and all subsequent calculations. 
The resulting mass-transfer timescale for Star~2 ($M_2/\dot{M}_2$) is $\sim1.26\times10^{7}~\mathrm{yr}$, comparable to its Kelvin--Helmholtz timescale ($\tau_{\rm KH,2} \sim GM_2^2/(R_2L_2) \sim1.40\times10^{7}~\mathrm{yr}$). 
This similarity indicates that the donor (Star~2) may deviate from strict thermal equilibrium.

For the periodic modulation, the Applegate mechanism \citep{1992ApJ...385..621A, 1998MNRAS.296..893L} can also be considered as an alternative to the LTT interpretation.
This mechanism involves variations in the stellar quadrupole moment driven by magnetic activity.
Assuming that the 18.376-yr periodic variation arises from the Applegate mechanism, we calculated the change in the gravitational quadrupole moment following \citet{1998MNRAS.296..893L}.
The resulting values are $2.824\times10^{49}\,\mathrm{g\,cm^{2}}$ for Star~1 and $7.101\times10^{49}\,\mathrm{g\,cm^{2}}$ for Star~2.
These values are approximately one to two orders of magnitude smaller than the typical range of $10^{51}$--$10^{52}$ reported for magnetically active close binaries such as RS~CVn by \citet{1998MNRAS.296..893L}.
Furthermore, as noted by \citet{2013AJ....145..100L}, the magnetic activity associated with the Applegate mechanism does not necessarily produce a smooth and curvilinear eclipse-timing variation pattern, which may limit its applicability to the BS~Cas system.
The observed eclipse-timing variations of BS~Cas are therefore most consistently explained by a continuous period decrease driven by AML and mass transfer between the components, superimposed on a periodic modulation produced by the LTT effect induced by a third body.

\subsection{Eclipse-depth Reversal} \label{sec:Results:eclipse-depth-revertial}

We compiled 13 contact binaries reported to exhibit eclipse-depth reversals, together with selected physical parameters, in Table~\ref{tab:subtype_reversal}. 
This table includes three systems (V410 Aur, RZ Com, and FG Hya) that exhibit eclipse-depth reversals among the eight contact binaries listed by \citet{2024ApJ...975..231X} as undergoing transitions between the A- and W-subtypes.
To maintain consistency with the notation used above, we designate the less massive component as Star~1.
In Table~\ref{tab:subtype_reversal}, parameter values presented as ranges rather than single values indicate that multiple solutions were obtained by analyzing different light curves for the same parameter.
When both cool- and hot-spot models were available, we preferentially adopted the cool-spot solutions.
The $dP/dt$ values for V410 Aur and AM Leo in the third column of Table~\ref{tab:subtype_reversal} were newly derived by the present authors using times of minima compiled in the TIDAK timing database, which includes recently observed minima.

Under the classification criterion of \citet{1984PASP...96..646B}, whereby systems are classified as A-subtype when the more massive component is hotter and as W-subtype otherwise, the systems listed in Table~\ref{tab:subtype_reversal} are generally classified as W-subtype.
However, several highly peculiar systems, such as V410 Aur, RZ Com, and FG Hya, exhibit classifications that alternate between W- and A-subtypes depending on the light curves adopted in the analysis.
Furthermore, strong O’Connell effects are apparent in the light curves of all systems, which can be explained by starspot activity (see the references in Table~\ref{tab:subtype_reversal} for details).
Most of the parameter values are widely distributed across broad ranges with no discernible trends: 0.225--0.440 days for $P$, 1.30--20.75 for $q$, 1.1--85.6\% for $f$, 69.1--89.860 $\degr$ for $i$, 4360--6439~K for $T_1$, and 4450--6307~K for $T_2$. 
However, the $dP/dt$ values indicate long-term period increases in nine systems, secular period decreases in two systems (BS~Cas and V2790~Ori), a sudden period change in TYC~4002-2628-1, and no significant change in CN~Tri.

As reported in Table~\ref{tab:subtype_reversal}, BS~Cas lies at the long-period and high-temperature end of the current sample of contact binaries reported to exhibit eclipse-depth reversals. 
In contrast to most of the other systems that undergo secular period increases, BS~Cas exhibits a large secular period decrease.
Accordingly, BS~Cas represents a valuable target for probing this phenomenon under relatively extreme system conditions and for examining its behavior in the context of the broader sample of eclipse-depth reversal systems.
At present, the sample of contact binaries exhibiting eclipse-depth reversals remains limited, and the system properties associated with this phenomenon are therefore poorly constrained.
Expanding the sample will therefore enable a population-level characterization of eclipse-depth reversals and provide a test of the possible explanations proposed by some investigators \citep{2024ApJ...975..231X, 2026ApJ...997..270W, 2026ApJS..282...17Z} for such apparent subtype conversions, thereby helping to better constrain the physical origin of this photometric behavior.

\subsection{Possible Origins of Surface-Brightness Inhomogeneity} \label{sec:discussion_spot_origin}
Asymmetries such as the O'Connell effect in the light curves of contact binaries are commonly attributed to either cool spots or processes associated with mass and energy transfer.
One traditional interpretation invokes a hot spot produced by the impact of transferred material on the stellar surface \citep[e.g.,][]{1997MNRAS.291..749H,2019PASJ...71...21K}.
In this scenario, the resulting surface-temperature inhomogeneity is expected to occur primarily on the mass-gaining component.
As mentioned in Section~\ref{sec:PeriodINV}, the orbital period of BS~Cas is gradually decreasing, which is consistent with mass transfer from the more massive Star~2 to the less massive Star~1 \citep{2001icbs.book.....H}.
Accordingly, a hot spot would be expected to be located on Star~1.
However, as discussed in Section~\ref{sec:Modeling:LCproperty}, the light variation is largest during the totality of the primary eclipse, when only Star~2 is visible.
This observational evidence implies that a mass-transfer-induced hot spot on Star~1 alone is insufficient to explain the observed long-term photometric variability of BS~Cas.

On the other hand, \citet{2009MNRAS.397..857S} treated the long-term mass transfer and the energy transport between the components as independent processes, with the latter carried by a flow from the hotter component toward the cooler component.
This distinction indicates that the mass transfer from Star~2 to Star~1 in BS~Cas does not necessarily preclude a separate lateral stream carrying mass and energy from the hotter Star~1 toward the cooler Star~2, as considered by \citet{2025ApJ...994....7F}.
However, because the lateral-transfer model of \citet{2025ApJ...994....7F} produces a steady-state longitudinal temperature gradient and an asymmetry between opposite quadratures, a persistent stream with the same flow direction would be expected to preferentially enhance the same quadrature. 
In contrast, Figure~\ref{fig:DifMag} shows reversals in the relative brightnesses of the two maxima of BS~Cas between epochs.
For comparison, we also tested phenomenological hot-spot configurations on both the hotter Star~1 (representing the traditional mass-transfer impact scenario) and the cooler Star~2 (motivated by the hotter-to-cooler energy-transfer scenario).
Both hot-spot configurations yielded slightly larger residuals than the cool-spot model listed in Table~\ref{tab:WDsolution}.
These considerations support the interpretation that, even if a stream-related surface-temperature enhancement is present on Star~2, such a stream alone appears insufficient to account for the observed long-term variations.
An additional time-variable cool-spot contribution is therefore still required in our interpretation, suggesting that cool-spot activity is likely the dominant source of the long-term photometric variability of BS~Cas.

The mass of Star~2 in BS~Cas is $M_2=1.423\pm0.031\,\rm{M_\odot}$, placing it in a regime where a deep convective envelope would not generally be expected for a single star. 
Additionally, as discussed by \citet{2025ApJ...995...19F}, energy transfer during the contact phase can further reduce the depth of the convective envelope.
However, the effective temperature of Star~2 ($T_{\rm eff}=6307~{\rm K}$) places it near the cool side of the Kraft break.\footnote{
The Kraft break is the sharp change in rotational velocities among main-sequence F-type stars \citep{1967ApJ...150..551K}, commonly attributed to the disappearance of the outer convective envelope at higher effective temperatures, rendering magnetic braking ineffective \citep{2024ApJ...973...28B}. \citet{2024ApJ...973...28B} quantified it at $T_{\rm eff}\simeq6550~{\rm K}$ with a width of about $200~{\rm K}$.}
Importantly, state-of-the-art 1-D stellar models \citep[e.g.,][]{2019ApJ...883..106C, 2022ApJS..262...19J} suggest that single stars in the mass range of $1.3$--$1.5\,M_\odot$ may still possess an active, highly turbulent, and compressible hydrogen-ionization convection zone (H I CZ) in their outermost layers.
Contact binaries are generally expected to have low Rossby numbers owing to their short orbital periods and synchronous rotation.
Observationally, they tend to occupy a lower-Rossby-number regime than single late-type stars \citep{1984A&A...133..117V, 1987ApJ...321..958V, 1993ASSL..177...51G}.
Accordingly, the fluid motion within this thin H I CZ is expected to be characterized by a low Rossby number, implying a strongly rotationally constrained convection regime \citep{2019ApJ...883..106C, 2022ApJS..262...19J}.
Under these conditions, dynamo action in the shallow convective layer may generate localized magnetic fields that can rise to the surface through magnetic buoyancy and produce magnetic surface features \citep{2019ApJ...883..106C}.

In addition, enhanced chromospheric or magnetic activity has been reported in close binaries relative to single stars \citep[e.g.,][]{1991A&A...251..183S, 2025NatAs...9.1045Y}.
More directly for contact binaries, \citet{2025ApJS..280...26S} detected flare activity in systems spanning ranges of effective temperature and stellar mass that overlap with the values inferred for Star~2 in BS~Cas (see their Figures 17 and 18).
Therefore, the presence of a shallow convective layer under strong rotational influence could enable Star~2 to develop magnetic cool spots. 
While the cool spot introduced in our WD modeling has a relatively large angular extent, it likely approximates the combined effect of multiple small magnetic cool spots, alongside complex brightness patterns unique to contact environments or local depressions caused by still poorly understood nonmagnetic mechanisms.

\subsection{Starspot Activity and Eclipse-Timing Variations}\label{sec:Results:ETVs_with_starspot}
Recently, quasi-periodic and anti-correlated variations between the primary and secondary times of minimum have been reported in many contact binaries with starspots (e.g., \citealt{2013ApJ...774...81T, 2015MNRAS.448..429B}); similar behavior may therefore be expected for BS~Cas.
To examine whether BS~Cas exhibits this behavior, we selected only the portion of the residual panel of Figure~\ref{fig:ETD} corresponding to the TESS observation and enlarged it, as shown in Figure~\ref{fig:ETD_TESS}.
This figure clearly demonstrates that BS~Cas exhibits quasi-periodic and anti-correlated timing variations. 
This behavior is generally interpreted as the migration of near-polar (i.e., low co-latitude) starspots induced by differential rotation on the stellar surface \citep{2013ApJ...774...81T, 2015MNRAS.448..429B}.
These properties suggest the presence of latitude-dependent differential rotation on the surface of BS~Cas.

Following \citet{2015MNRAS.448..429B}, we define the ratio between the orbital and spot-migration periods as $|\kappa| \equiv P_{\mathrm{orb}}/P_{\mathrm{mig}}$, where $P_{mig}$ is the characteristic starspot-migration period.
From Figure~\ref{fig:ETD_TESS}, we estimate the migration period of BS~Cas to be approximately twice the duration of a single TESS sector (i.e., $\sim55$~days).
We then obtain $|\kappa| \sim 0.008$, which lies near the upper end of the $|\kappa|$ distribution shown in \citet{2015MNRAS.448..429B}. 
This value exceeds those of the majority of contact binaries in the sample of \citet{2015MNRAS.448..429B} and remains within the range expected for single stars.
In addition, $T_2 = 6307$~K lies near the peak of the effective-temperature distribution presented by \citet{2015MNRAS.448..429B} for contact binaries that exhibit anti-correlated variations.

The amplitude of the anti-correlated variation observed in BS~Cas is approximately $\pm50$~s, which is significantly smaller than the $\pm 200$--$300$~s reported by \citet{2013ApJ...774...81T}.
\citet{2013ApJ...774...81T} also suggested that the anti-correlation model is particularly applicable to systems with lower orbital inclination. 
They argued that, in such systems, the starspot on the more massive component is not obscured during eclipse and thus remains continuously visible.
The relatively small amplitude observed in BS~Cas is therefore likely attributable to its nearly edge-on inclination ($i\approx 90\degr$).

Based on the simulation by \citet{2013ApJ...774...81T}, which predicts a symmetric anti-correlated variation pattern, we expect the epoch-averaged residuals of the primary and secondary minima to remain approximately constant.
However, Figure~\ref{fig:ETD_TESS} shows that the averaged residuals vary with time, with the largest deviation occurring in Sector~58.
Moreover, the averaged residuals exhibit an overall decreasing trend of $\sim 80$~s.
These observational features suggest that additional physical effects cannot be excluded, such as an LTT variation ($\tau_4$) induced by a fourth body, the Applegate mechanism, and in-phase modulations of the primary and secondary eclipse timings that may arise from starspots \citep{2013ApJ...774...81T}.
Determining the causes of these variations will require long-term and high-precision measurements of eclipse timings.

\subsection{Constraints on the nature of the additional object in BS~Cas}\label{sec:Results:ETV}
In the previous sections, we identified three observational signatures that may be related to an additional object in the BS~Cas system.
Specifically, the 2020 spectroscopic observations reveal a third peak in the BF profiles, the light-curve modeling yields a statistically significant third-light contribution, $\ell_3$, and the ETD shows a periodic modulation of the orbital period.
In this subsection, we compare these three signatures to assess their mutual consistency and to explore the nature of the additional object.

\begin{figure}
\epsscale{0.8}
\plotone{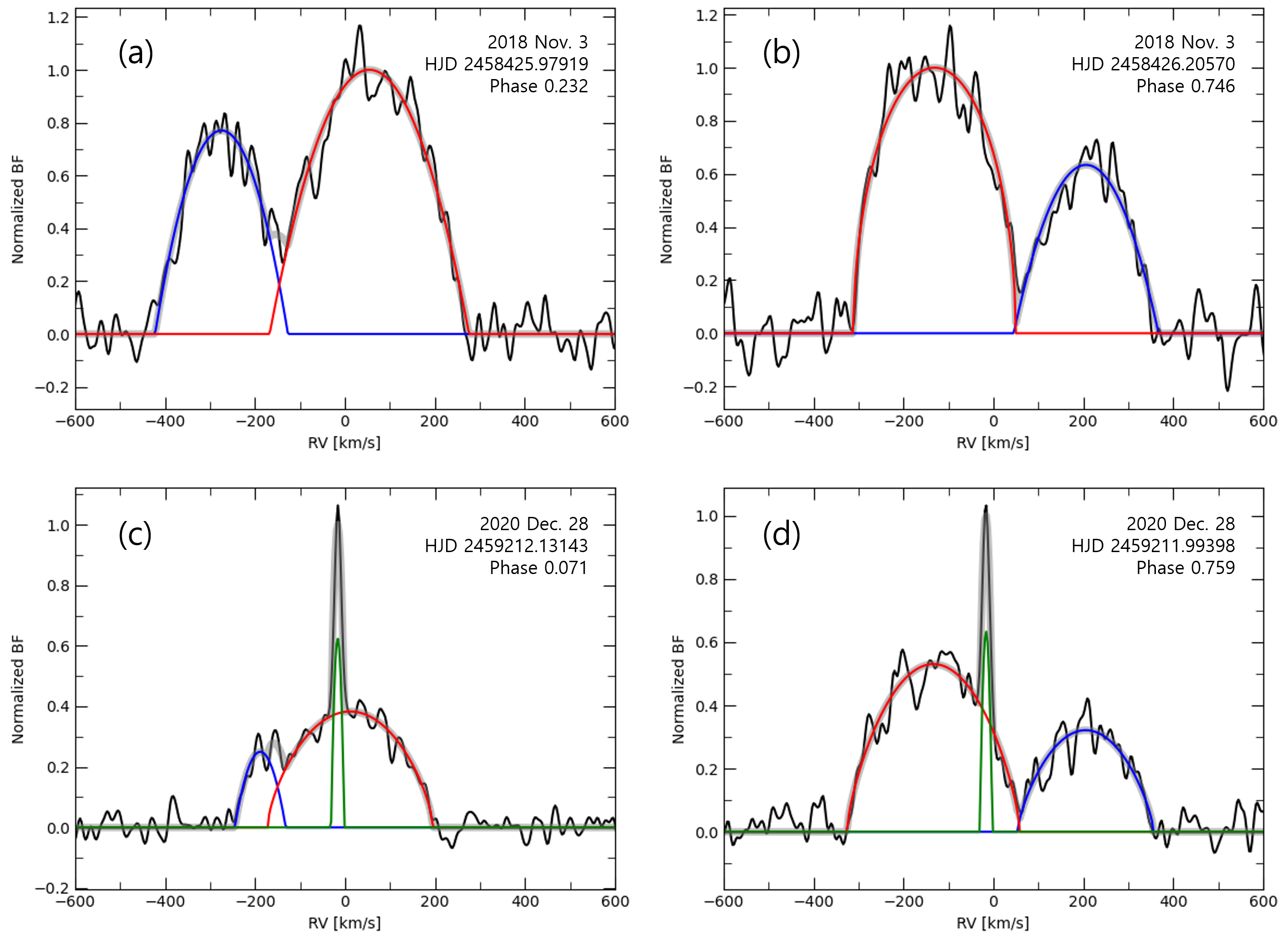}
\caption{Samples of BF profiles derived with \texttt{RaveSpan}.
Panels (a)--(d) show the BF profiles obtained on 2018 November 3 at phases 0.232 (HJD 2458425.97919) and 0.746 (HJD 2458426.20570), and on 2020 December 28 at phases 0.071 (HJD 2459212.13143) and 0.759 (HJD 2459211.99398), respectively.
The black lines indicate the BF profiles derived with \texttt{RaveSpan}.
The blue, red, and green lines denote the individual rotationally broadened BF components associated with Star~1, Star~2, and Star~3, respectively.
The light gray lines represent the combined profiles of the three fitted components.
}
\label{fig:BF}
\end{figure}

\begin{figure}
\epsscale{0.8}
\plotone{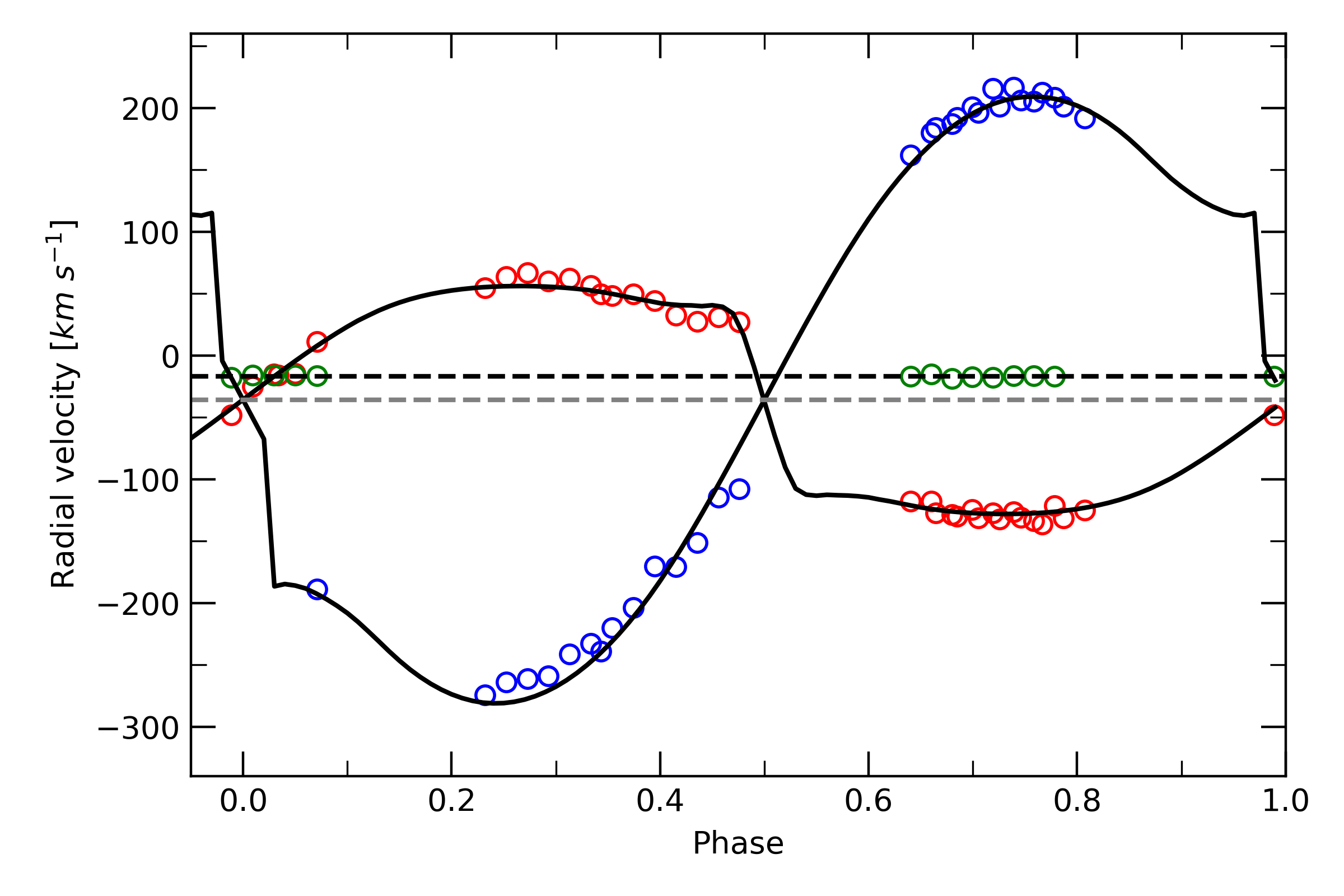}
\caption{RV curves of BS~Cas. The blue, red, and green circles represent Star~1 ($v_1$), Star~2 ($v_2$), and the third component ($v_3$), respectively. 
The black solid curves represent the results of the simultaneous solution. 
The black dashed line indicates the mean RV of the third component. 
The gray dashed line corresponds to the systemic velocity ($\gamma_{0}$) of the contact binary system derived from the WD solution.
}
\label{fig:RVf}
\end{figure}

\begin{figure}
\epsscale{1.0}
\plotone{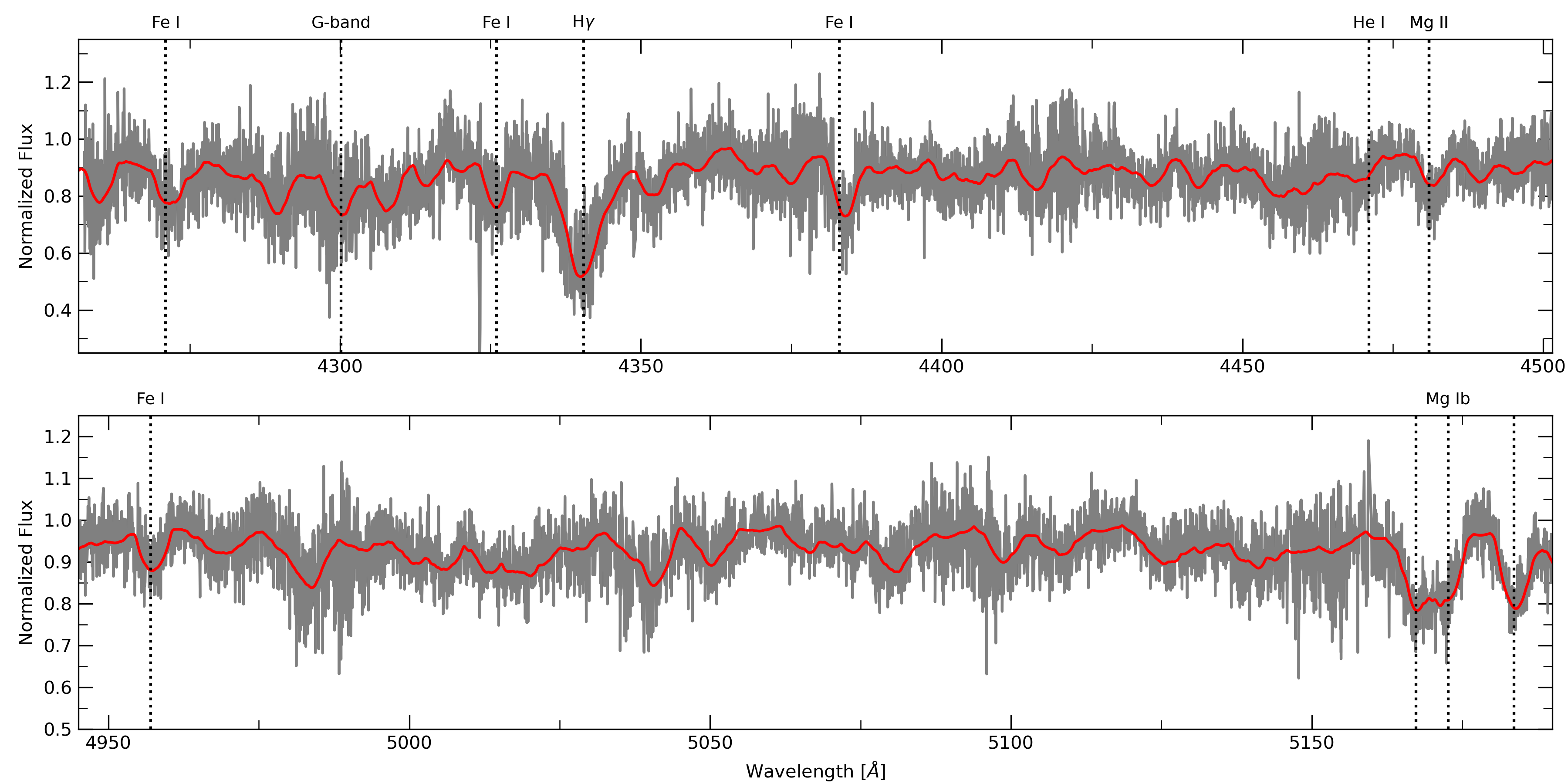}
\caption{Normalized spectrum of BS~Cas covering the G band and the Mg I b region obtained at HJD 2458427.2132.
The gray line represents the observed spectrum.
The red line denotes the synthetic spectrum with $T_2 = 6307~\mathrm{K}$ and $v_2\sin i = 176~\mathrm{km\,s^{-1}}$.}
\label{fig:Spec_Results}
\end{figure}

\begin{figure}
\epsscale{0.7}
\plotone{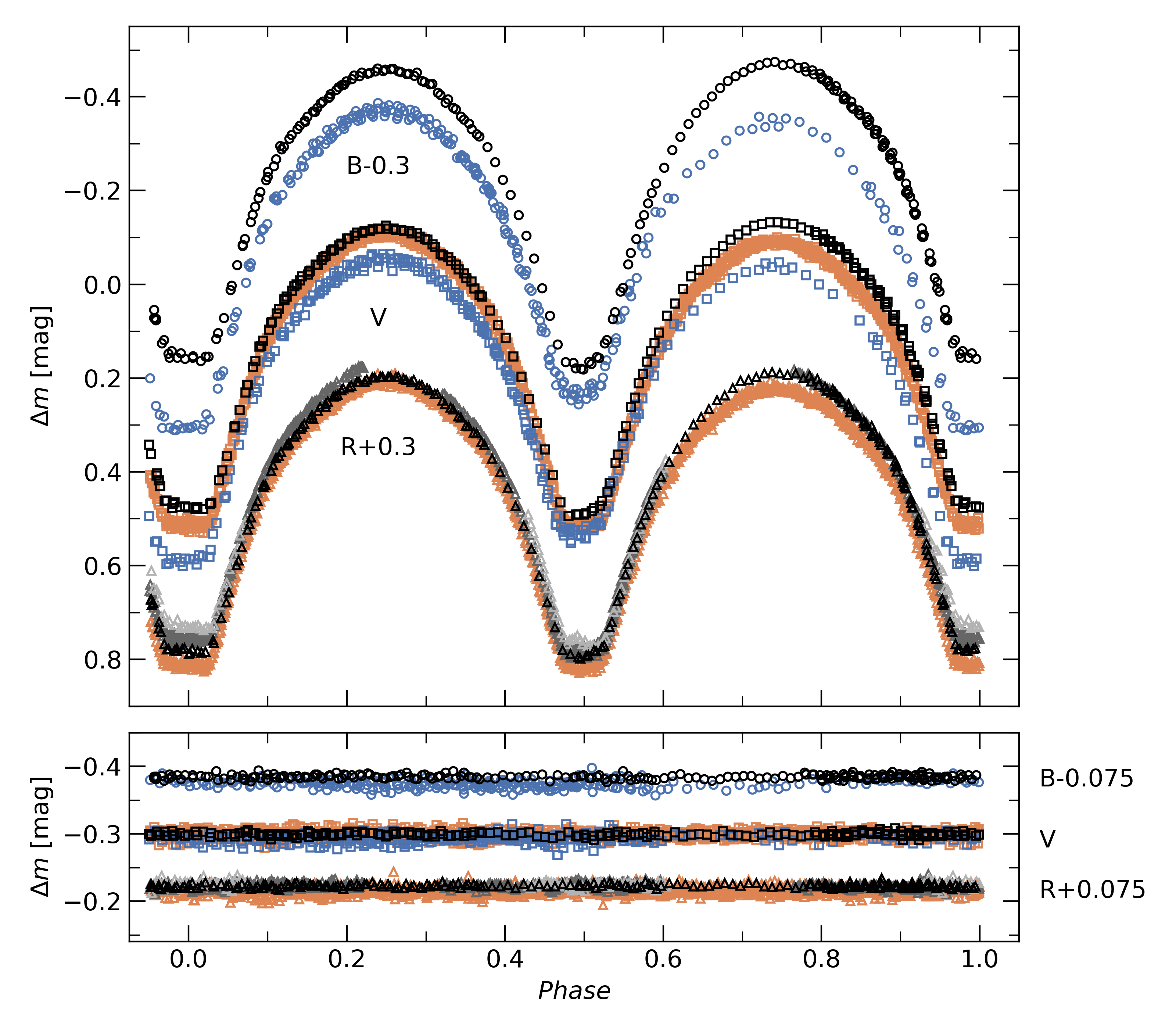}
\caption{Phase-folded $BVR$ light curves of BS~Cas obtained in 2011 at CbNUOJ (blue symbols), 
in 2014--2015, 2018, and 2020 at SOAO (dark gray, light gray, and black), and in 2024 at LOAO (orange). 
The upper panel shows all differential light curves, and the lower panel shows the magnitude differences between the comparison and check stars, vertically shifted for clarity. 
The circles, squares, and triangles denote the $B$, $V$, and $R$ bands, respectively. 
In the upper (lower) panel, the $B$ and $R$ curves are offset by $-0.30$ ($-0.075$) and 
$+0.30$ ($+0.075$) mag relative to the $V$ curves.}
\label{fig:LCour}
\end{figure}

The 2020 BF profiles in Figure~\ref{fig:BF} clearly reveal the presence of an additional object ($S_{\rm BF}$) located along the same line of sight as the contact binary.
The light contribution of $S_{\rm BF}$, estimated from the area ratio of the fitted BF components, is 3.4--7.5\% of the total system light.
The BF profiles were derived from the 4700--5450\,\AA\ spectral region, which overlaps with the long-wavelength end of the $B$ band and the short-wavelength portion of the $V$ band.
We therefore compared the light contribution of $S_{\rm BF}$ with the $\ell_{3,B}$ and $\ell_{3,V}$ values derived from the 2020 SOAO light curves; the results are consistent within the uncertainties.

We next examined the properties of the LTT companion ($S_{\rm LTT}$) inferred from the orbital-period analysis and compared them with those of $S_{\rm BF}$ and $\ell_3$.
Assuming that $S_{\rm LTT}$ is gravitationally bound and responsible for the observed LTT variation, we derived the minimum mass of the LTT companion as $M_{3,i_3=90} = 0.24\,\mathrm{M_\odot}$ from the mass function using Tables~\ref{tab:Period_Inv_Results} and \ref{tab:WDsolution}. 
If the third body is a main-sequence star, its spectral type can be estimated to be roughly M3 V using the observational main-sequence parameter table of \citet{2000PhT....53j..77C}.
The corresponding fractional luminosity is $\ell_3\sim0.076\%$, which is smaller than the $\sim 1\%$ practical detection limit for $\ell_3$ reported by \citet{2021PASP..133h4202L} and is therefore effectively undetectable in the WD modeling.
Alternatively, if $S_{\rm LTT}$ contributes approximately 10.5\% of the total system light, i.e., the maximum $\ell_3$ value in Table~\ref{tab:WDsolution_All}, its mass could be as large as $0.86\,\mathrm{M_\odot}$ for an orbital inclination of $i_3 \approx 19.5\degr$.
In this case, a main-sequence $S_{\rm LTT}$ would have a mass in the range $0.24$--$0.86\,\mathrm{M_\odot}$. 

In terms of their intrinsic variability, main-sequence stars in this mass range have relatively stable internal energy transport structures and are not expected to undergo large secular luminosity changes.
Nevertheless, rapidly rotating late-type main-sequence stars can exhibit strong starspot activity and may therefore display substantial photometric variability.
Even with extreme starspot activity, however, variability of $S_{\rm LTT}$ alone is unlikely to produce system-wide brightness changes at the $\sim 10\%$ level.
For this reason, it would be difficult for such a star to account for either the third peak observed in the 2020 BF profiles only (Figure~\ref{fig:BF}) or the long-term brightness variations shown in Figure~\ref{fig:DifMag}(c).
A similar argument applies if $S_{\rm LTT}$ is a white dwarf or a brown dwarf.

To examine whether $S_{\rm LTT}$ is the same object as $S_{\rm BF}$, we computed the expected RV curves of $S_{\rm LTT}$ and the contact binary using the ETV solutions listed in Table~\ref{tab:Period_Inv_Results}, as shown in Figure~\ref{fig:RV_m12_m3}.
Using the systemic velocity $\gamma_0 = -35.97\, \mathrm{km\,s^{-1}}$, obtained from a simultaneous analysis of the 2018--2020 RV and light-curve data, we adopted $\gamma_{12,3}=-35.50\, \mathrm{km\,s^{-1}}$ for the combined system consisting of the contact binary and $S_{\rm LTT}$.
For a given mass function, the mass $M_3$ depends on the orbital inclination $i_3$, which determines the semi-amplitude $K_3$ of the predicted RV curve.
As shown in Figure \ref{fig:RV_m12_m3}, the maximum $K_3$ is obtained in the most favorable case of minimum mass $M_{3,i_3=90}$.
Even in this most favorable scenario, the predicted RV curve of $S_{\rm LTT}$ does not reproduce the observed RV $v_3$ of $S_{\rm BF}$.
We therefore conclude that $S_{\rm LTT}$ is highly unlikely to be the same object as $S_{\rm BF}$. 
Thus, while $S_{\rm LTT}$ is interpreted as a gravitationally bound companion responsible for the observed LTT signal, $S_{\rm BF}$ is likely a separate unidentified object projected along the same line of sight as BS~Cas and not bound to the system.

\section{Summary and Conclusions}\label{sec:Summary}
\begin{figure}
\epsscale{0.54}
\plotone{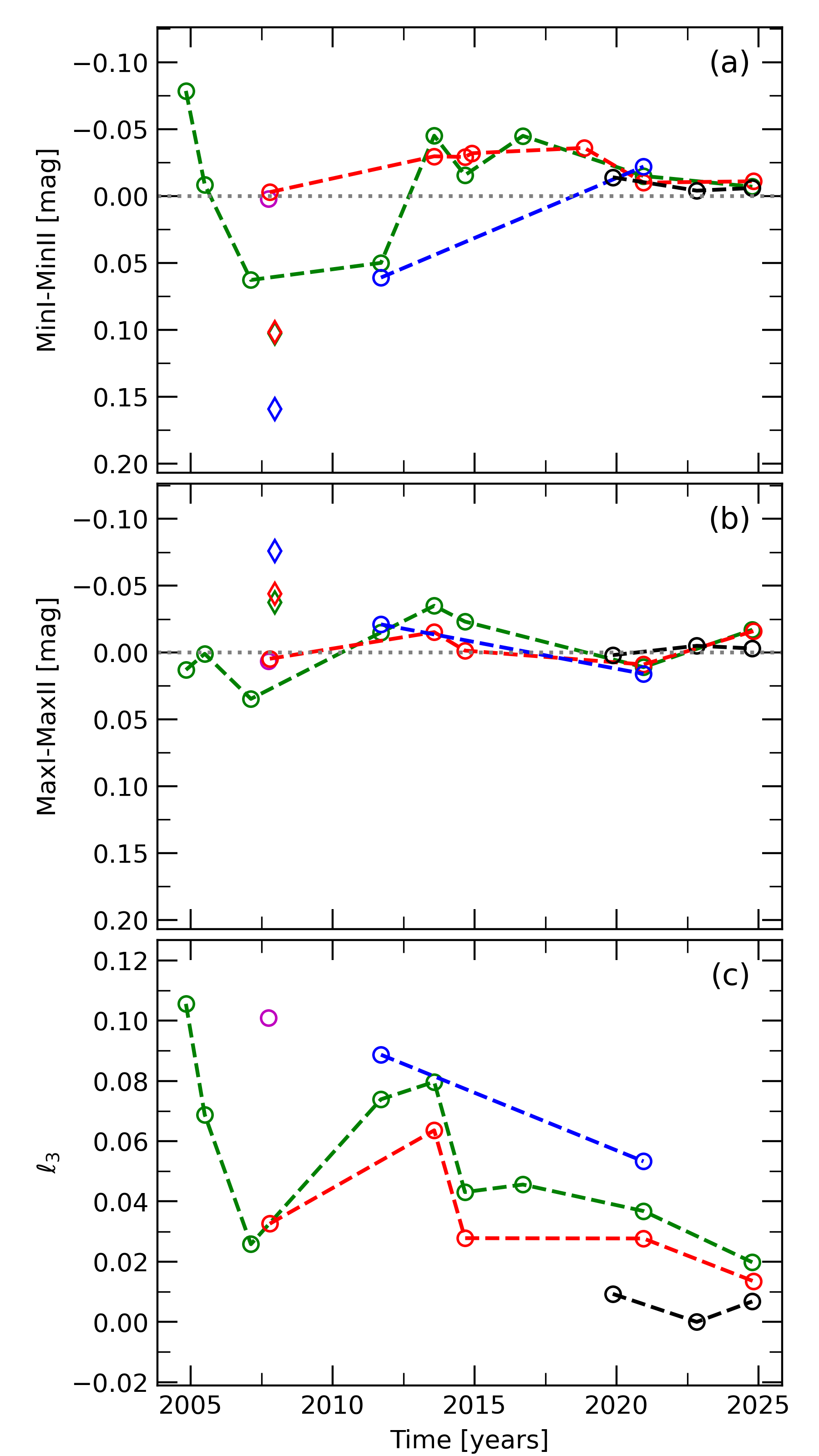}
\caption{Variations of differential light, $\mathrm{Min~I-Min~II}$ (a) and $\mathrm{Max~I-Max~II}$ (b), and $\ell_3$ (c) in BS~Cas. 
The blue, green, red, magenta, and black symbols indicate $B$, $V$, $R$, $wide$, and $TESS$ filters, respectively. 
The diamonds denote the data adopted from Y08.}
\label{fig:DifMag}
\end{figure}

\begin{figure}
\epsscale{0.5}
\plotone{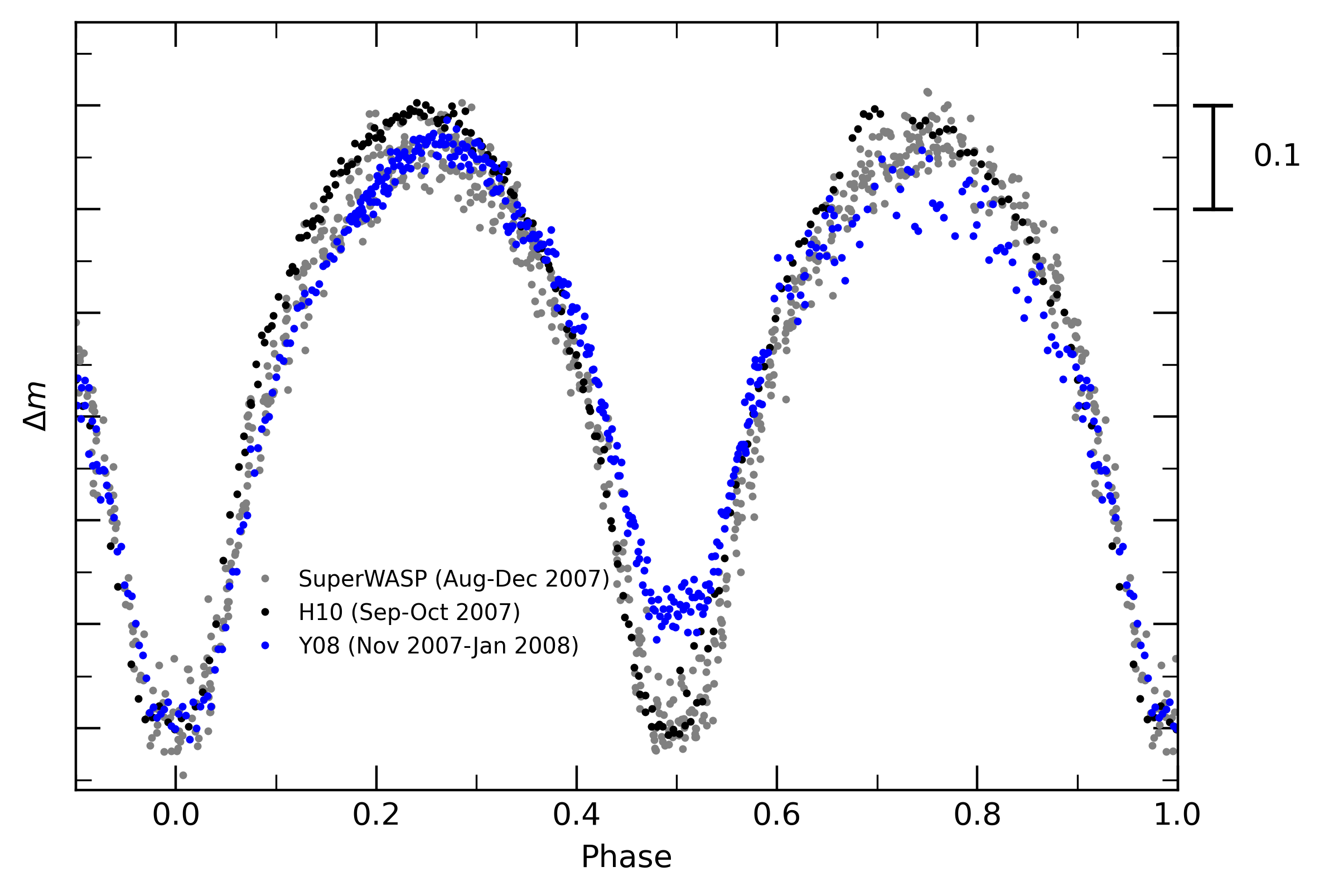}
\caption{
Comparison of the Y08 ($R$ band; blue dots), H10 ($R$ band; black dots), and SuperWASP (wide band; gray dots) light curves obtained during the same observing interval. The light curves were vertically shifted to match their levels at $\mathrm{Min\,I}$. The vertical scale bar represents 0.1 mag.}
\label{fig:Y08H10SWASP}
\end{figure}

\begin{figure}
\epsscale{1.0}
\plotone{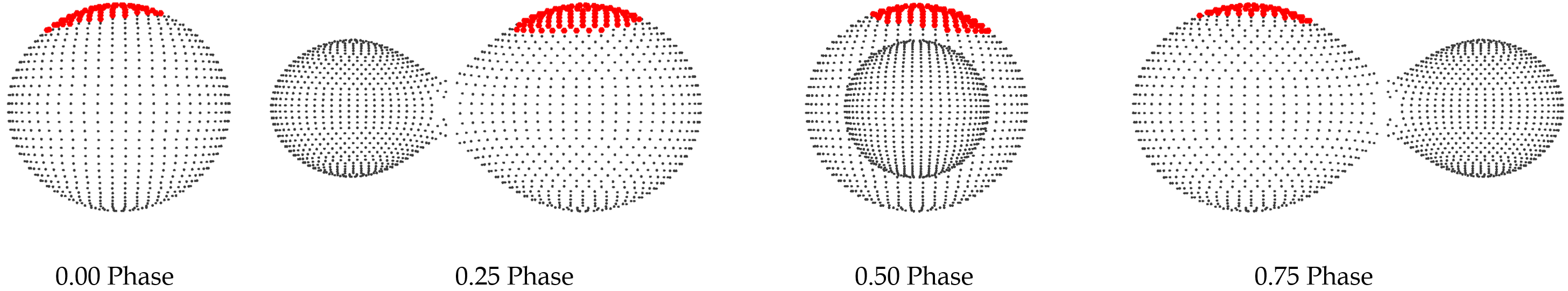}
\caption{
Surface configurations of BS~Cas for the adopted Star~2 cool-spot model based on the 2020 SOAO $BVR$ light-curve solution at four representative orbital phases. Red points indicate the cool-spot region.
}
\label{fig:3D_config}
\end{figure}

\begin{figure}
\epsscale{1.0}
\plotone{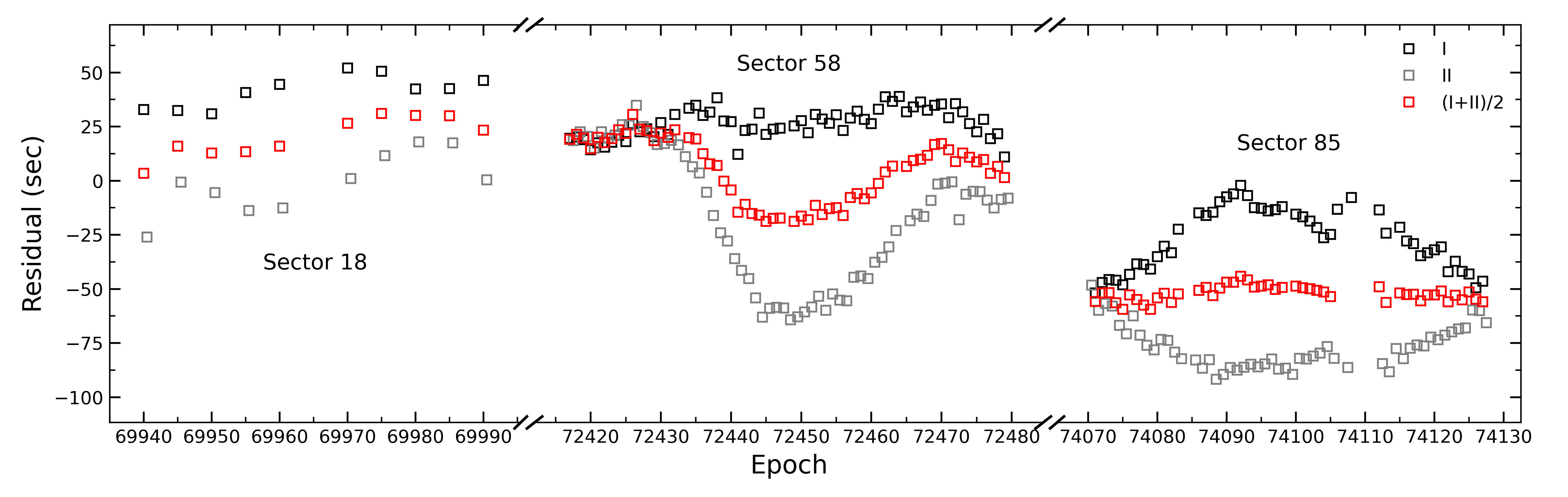}
\caption{TESS eclipse-timing residuals for BS~Cas. 
The black and gray symbols are identical to those used in Figure \ref{fig:ETD}.
The red squares represent the epoch-averaged residuals for the primary and secondary minima.}
\label{fig:ETD_TESS}
\end{figure}

\begin{figure}
\epsscale{0.7}
\plotone{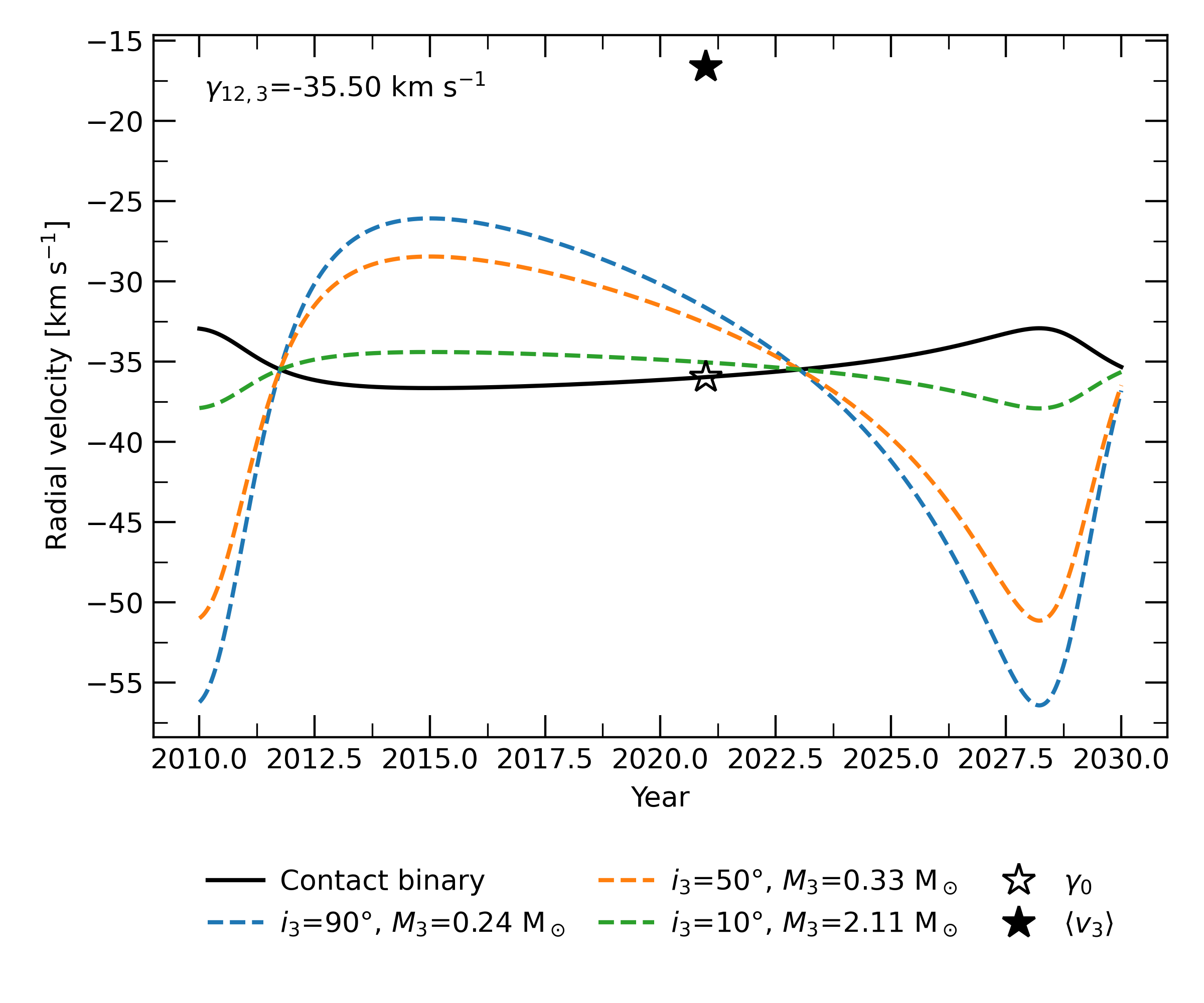}
\caption{RV curves reconstructed from our analysis results.
The solid and dashed lines represent the RV curves of the contact binary system and $M_3$, respectively.
The colors denote the assumed outer-orbit inclination $i_3$ (blue: $90\degr$; orange: $50\degr$; green: $10\degr$).
The open and filled stars indicate $\gamma_{0}$ derived from the WD modeling and the mean $v_3$ value, respectively.
}
\label{fig:RV_m12_m3}
\end{figure}

We conducted a comprehensive study of BS~Cas using 22~yr of multi-band photometric data from several observatories, together with 2~yr of high-resolution spectroscopic observations.
The main results are summarized as follows:

\begin{itemize}
\item[1.] The historical light curves spanning 22~yr exhibit significant variations, particularly in the depth of the total eclipse.
Notably, the overall brightness level in 2011 was lower than that observed in 2020. 
The variation patterns of $\mathrm{Min~I-Min~II}$ and $\mathrm{Max~I-Max~II}$ resemble segments of a sinusoidal modulation.
As a result of these variations, the light-curve morphology of BS Cas appears to alternate between the A- and W-subtypes.
\item[2.] The first double-lined RV curves were derived from the BF profiles. 
An additional third component appears only in the 2020 BF profiles and is absent in the 2018 data.
This spectroscopic component is not expected to be gravitationally bound to the BS~Cas system.
\item[3.] From the simultaneous analysis of the RV and light curves, we derived $T_1=6439$ K, $T_2=6307$ K, $i=89.860\pm0.175\degr$, $f = 32.25 \%$, and $q=2.515\pm0.003$.
The absolute parameters are estimated as $M_1=0.566\pm0.011~\rm{M_\odot}$, $M_{2}=1.423\pm0.031~\rm{M_\odot}$, $R_1=0.986\pm0.007~\rm{R_\odot}$, $R_2=1.473\pm0.010~\rm{R_\odot}$, $L_1=1.507\pm0.078~\rm{L_\odot}$, and $L_2=3.093\pm0.149~\rm{L_\odot}$. 
These parameters indicate that BS~Cas is a W-subtype contact binary system comprising a cooler, more massive component and a hotter, less massive component.
\item[4.] The orbital-period analysis reveals a continuous decrease at a rate of $dP/dt=-2.305\pm0.004\times 10^{-7}~\mathrm{days~yr^{-1}}$.
This decrease is consistent with mass transfer from Star~2 to Star~1 combined with AML.
In addition, the period of the cyclic modulation was estimated as $P_3 = 18.376 \pm 0.076$~yr.
This variation is likely caused by the gravitational influence of a third body bound to the contact binary system rather than by the Applegate mechanism.
\item[5.] Pronounced changes are observed in the light-curve morphology (including the O’Connell effect and eclipse depth), together with clear anti-correlated variations between the times of primary and secondary minima. 
These behaviors are likely driven by the evolution of multiple starspots located at different co-latitudes on the surface of the secondary star undergoing differential rotation. 
In particular, when systems exhibiting eclipse-depth reversal are analyzed solely on the basis of photometric data, they may be misinterpreted as systems undergoing a transition from the A-subtype to the W-subtype, or vice versa, on short timescales. 
In such cases, RV curves derived from spectroscopic observations can resolve this ambiguity.
\item[6.] The object $S_{\rm LTT}$, which produces the observed LTT signal, appears to be a gravitationally bound and relatively stable companion.
By contrast, the object $S_{\rm BF}$ responsible for the third peak observed in the 2020 BF profiles is likely associated with a wandering object that is not gravitationally bound to the BS~Cas system, but happens to move along the same line of sight as BS~Cas. 
In addition, it remains unclear whether the variation in the third light shown in Figure \ref{fig:DifMag}(c) is related to $S_{\rm LTT}$, $S_{\rm BF}$, or another unknown cause.
\end{itemize} 

BS~Cas provides a representative case that highlights the value of long-term photometric and spectroscopic monitoring in probing the complex physical processes operating in contact binary systems.  
At present, however, the available observational material does not allow a definitive identification of the nature of $S_{\rm BF}$, nor does it permit stringent constraints on its physical properties. 
In addition, the current dataset is insufficient for a detailed investigation of the surface spot activity and latitude-dependent differential rotation of the contact-binary components.
We therefore anticipate that, when combined with continued long-term photometric monitoring, future high-resolution, high-signal-to-noise spectroscopic observations will enable deeper insight into the magnetic activity, spot evolution, and mutual interactions between the components of BS~Cas.

\par
\vspace{1em}
\noindent
We would like to express our sincere gratitude to the anonymous reviewer, whose thorough review and encouraging comments enhanced this manuscript and clarified key points.
We also appreciate the efforts of all staff at the Chungbuk National University Observatory (Jincheon Station), which is operated by Chungbuk National University, and at the Bohyunsan, Sobaeksan, and Lemmonsan Optical Astronomy Observatories operated by the Korea Astronomy and Space Science Institute (KASI).
This paper makes use of data collected with the TESS mission, obtained from the MAST data archive at the Space Telescope Science Institute (STScI). 
This research made use of TESS Full Frame Images (FFIs) from Sectors 18, 58, and 85, accessed via the TESScut service (\dataset[doi:10.17909/0cp4-2j79]{https://doi.org/10.17909/0cp4-2j79}).
Funding for the TESS mission is provided by the NASA Explorer Program. 
STScI is operated by the Association of Universities for Research in Astronomy, Inc., under NASA contract NAS 5-26555.
This research made use of Lightkurve, a Python package for Kepler and TESS data analysis \citep{2018ascl.soft12013L}.
We also used public data from the DR1 release of the WASP survey \citep{2010A&A...520L..10B}, as provided by the WASP consortium, and computational resources supplied by the project ``e-Infrastruktura CZ'' (e-INFRA CZ LM2018140), supported by the Ministry of Education, Youth and Sports of the Czech Republic.
Additional data were obtained from the OMC Archive at CAB (INTA-CSIC), pre-processed by ISDC and further processed by the OMC Team at CAB.
The OMC Archive is part of the Spanish Virtual Observatory project. Both are funded by MCIN/AEI/10.13039/501100011033 through grants PID2020-112949GB-I00 and PID2019-107061GB-C61, respectively.
In addition, we acknowledge with thanks the variable star observations from the AAVSO International Database contributed by observers worldwide and used in this research.
This work has also made use of data from the European Space Agency (ESA) mission {\it Gaia} (\url{https://www.cosmos.esa.int/gaia}), processed by the {\it Gaia} Data Processing and Analysis Consortium (DPAC, \url{https://www.cosmos.esa.int/web/gaia/dpac/consortium}). 
Funding for the DPAC has been provided by national institutions, in particular the institutions participating in the {\it Gaia} Multilateral Agreement.
This research has also made use of the SIMBAD database, operated at CDS, Strasbourg Astronomical Observatory, France.
This work made use of Astropy: a community-developed core Python package and an ecosystem of tools and resources for astronomy \citep{astropy:2013, astropy:2018, astropy:2022}.
Finally, this study was supported by the National Research Foundation of Korea (NRF; grant numbers RS-2021-NR066086, RS-2024-00452238, and RS-2025-21073000) and by the Korea Astronomy and Space Science Institute (KASI; project number 2026-1-904-01).

\bibliography{new.ms}{}
\bibliographystyle{aasjournal}

\begin{deluxetable}{cccc}
\tabletypesize{\scriptsize}
\tablewidth{0pt}
\tablecaption{Sample of the Photometric Data for BS~Cas.}
\small
\tablehead{HJD & \textit{$\Delta mag$} & Band & Observatory}
\startdata
2455827.28996  & 0.030  & $B$  & CbNUOJ  \\
2455827.29156  & 0.053  & $V$  & CbNUOJ  \\
2455827.29317  & 0.040  & $B$  & CbNUOJ  \\
2455827.29478  & 0.063  & $V$  & CbNUOJ  \\
2455827.29724  & 0.054  & $B$  & CbNUOJ  \\
2455827.30029  & 0.092  & $V$  & CbNUOJ  \\
.. & .. & .. & .. \\
\enddata
\tablecomments{All measurements obtained at CbNUOJ, SOAO, and LOAO are offered online.}
\label{tab:LCdata}
\end{deluxetable}

\begin{deluxetable}{lcccc}
\tabletypesize{\scriptsize}
\tablewidth{0pt}
\tablecaption{Summary of Information on the Photometric Data.}
\small
\tablehead{
\colhead{Observatory} &
\multicolumn{2}{c}{All data} &
\multicolumn{2}{c}{Selected dataset for LC} \\
\cline{2-3}\cline{4-5}
\colhead{} &
\colhead{Observational Date} &
\colhead{Filter/No.} &
\colhead{Observational Date} &
\colhead{Filter/No.}
}
\startdata
OMC       & Mar 2003--Jul 2023         & $V$/2943               & Sep--Dec 2004            & $V$/297         \\
          &                               &                      & Jun--Jul 2005               & $V$/930         \\
          &                               &                      & Feb 2007                 & $V$/230         \\
          &                               &                      & Dec 2020--Feb 2021            & $V$/427         \\[1pt]
\noalign{\vskip 4.5pt}
SuperWASP & Aug 2007--Jul 2008        & $wide$/1176            & Aug--Dec 2007         & $wide$/822      \\[1pt]
\noalign{\vskip 4.5pt}
H10       & Sep--Oct 2007            & $R$/221                & Sep--Oct 2007            & $R$/221         \\[1pt]
\noalign{\vskip 4.5pt}
Y08       & Nov 2007--Jan 2008            & $B$/428; $V$/455; $R$/477  & -                            & -            \\[1pt]
\noalign{\vskip 4.5pt}
AAVSO     & Jul 2012--Jan 2020            & $V$/6346; $R$/5720; $I$/55 & Jul--Aug 2013            & $V$/2920; $R$/2892\\
          &                               &                      & Aug--Sep 2014            & $V$/2828; $R$/2849\\
          &                               &                      & Sep 2016            & $V$/152         \\[1pt]
\noalign{\vskip 4.5pt}
TESS      & Nov 2019 (Sector 18)          & $TESS$/1103          & November 2019            & $TESS$/1028     \\
          & Oct--Nov 2022 (Sector 58)     & $TESS$/11692         & Oct--Nov 2022            & $TESS$/11692    \\
          & Oct--Nov 2024 (Sector 85)     & $TESS$/10586         & Oct--Nov 2024            & $TESS$/9942     \\[1pt]
\noalign{\vskip 4.5pt}
CbNUOJ    & Sep 2011                      & $B$/269; $V$/263         & Sep 2011                  & $B$/269; $V$/263  \\[1pt]
\noalign{\vskip 4.5pt}
SOAO      & Sep 2014--Dec 2020            & $B$/211; $V$/210; $R$/869  & Dec 2020                      & $B$/211; $V$/210; $R$/215\\[1pt]
\noalign{\vskip 4.5pt}
LOAO      & October 2024                  & $V$/1034; $R$/1037       & Oct 2024                  & $V$/1034; $R$/1037  \\
\enddata
\tablecomments{`No.' indicates the number of the data points.}
\label{tab:Info_PhotoData}
\end{deluxetable}

\clearpage

\startlongtable
\begin{deluxetable*}{cccccc}
\tablecaption{Sample of the Observed Times of Minimum Lights.}
\tabletypesize{\scriptsize}
\tablehead{
\colhead{HJD} & \colhead{Error} & \colhead{Epoch} & \colhead{Method${}^{\rm{*}}$} & \colhead{Type} & \colhead{Reference}
}
\startdata
2427984.469  &              & -0.5    & P   & II & \citet{1947VeSon...1...43A} \\
2428343.462  &              & 814.5   & P   & II & \citet{1947VeSon...1...43A} \\
2428904.228  &              & 2087.5  & P   & II & \citet{1947VeSon...1...43A} \\
2429106.416  &              & 2546.5  & P   & II & \citet{1947VeSon...1...43A} \\
2429114.538  &              & 2565    & P   & I  & \citet{1947VeSon...1...43A} \\
\nodata & \nodata & \nodata & \nodata & \nodata & \nodata  \\
\enddata
\tablenotetext{\rm{*}}{P: Plate, PG: Photographic, CCD: charge-coupled device}
\tablerefs{
\citet{1947VeSon...1...43A}, \citet{1992BAVSM..60....1H}, \citet{2005IBVS.5643....1H}, \citet{1997BBSAG..114..5}, \citet{1998BBSAG.117....9D}, \citet{2001BBSAG..125..3},   \citet{2003IBVS.5378....1D}, \citet{2005IBVS.5592....1K}, \citet{2005IBVS.5668....1P}, \citet{2005IBVS.5657....1H}, \citet{2007IBVS.5809....1S}, \citet{2006IBVS.5677....1D}, \citet{2007IBVS.5761....1H}, \citet{2006IBVS.5731....1H}, \citet{2007IBVS.5777....1P}, \citet{2007IBVS.5795....1D}, \citet{2008IBVS.5830....1H}, \citet{2007OEJV...74....1B}, \citet{2010RAA....10..569H}, \citet{2008AJ....136..594Y}, \citet{2009IBVS.5889....1H}, \citet{2009IBVS.5898....1P}, \citet{2008OEJV...94....1B}, \citet{2009OEJV..107....1B}, \citet{2009IBVS.5871....1D}, \citet{2011OEJV..137....1B}, \citet{2010IBVS.5941....1H}, \citet{2011IBVS.5980....1P}, \citet{2011IBVS.5984....1H}, \citet{2011IBVS.5966....1N}, \citet{2012IBVS.6011....1D}, \citet{2012IBVS.6018....1N}, \citet{2012IBVS.6026....1H}, \citet{2013IBVS.6048....1H}, \citet{2013IBVS.6042....1D}, \citet{2014IBVS.6118....1H}, \citet{2014IBVS.6092....1N}, \citet{2013OEJV..160....1H}, \citet{2015IBVS.6152....1H}, \citet{2016IBVS.6167....1P}, \citet{2017OEJV..179....1J}, \citet{2017IBVS.6196....1H}, \citet{2017IBVS.6195....1N}, \citet{2021OEJV..211....1L}, \citet{2018IBVS.6244....1P}, \citet{2019BAVJ...31....1P}, \citet{2020JAVSO..48..169N}, \citet{2022BAVJ...60....1P}, \citet{2023JAVSO..51..134S}, VarAstro (Bragagnolo, U.),   VarAstro (Ehrenberger, R.),  VarAstro (Jacobsen, J.) 
}
\label{tab:MINdata}
\end{deluxetable*}

\clearpage

\begin{deluxetable}{ccc}
\tabletypesize{\scriptsize}
\tablewidth{0pt}
\tablecaption{Orbital parameters for the LTT effect.}
\label{tab:Period_Inv_Results}
\tablehead{Parameter & Value & Unit }
\startdata
$T_0$              & $2,427,984.70958 \pm0.00094$   & HJD  \\
$P$                & $0.440483721 \pm0.000000029$   & day  \\
$a_{12}\sin{i_3}$  & $0.996 \pm0.028$        & AU          \\
$\omega_{12}$      & $40.650 \pm3.467$       & $\degr$    \\
$e_3$              & $0.500 \pm0.032$        &             \\
$T_{12}$           & $2,428,525 \pm102$      & HJD         \\
$P_3$              & $18.376 \pm0.076$       & year        \\
$K$                & $0.00532\pm0.00017$     & day         \\
$A$                & $(-1.390 \pm0.002) \times 10^{-10}$  & day  \\
$dP/dt$            & $(-2.305 \pm0.004) \times 10^{-7}$  & days~yr$^{-1}$ \\
\enddata
\end{deluxetable}

\begin{deluxetable}{ccccccc}
\tabletypesize{\scriptsize}
\tablewidth{0pt}
\label{tab:RVt}
\tablecaption{Radial velocities of BS~Cas.}
\tablehead{
\colhead{\begin{tabular}{@{}c@{}}HJD\\(+2,450,000)\end{tabular}} &
\colhead{\begin{tabular}{@{}c@{}}$v_1$\\$[\rm km\,s^{-1}]$\end{tabular}} &
\colhead{\begin{tabular}{@{}c@{}}$\sigma_1$\\$[\rm km\,s^{-1}]$\end{tabular}} &
\colhead{\begin{tabular}{@{}c@{}}$v_2$\\$[\rm km\,s^{-1}]$\end{tabular}} &
\colhead{\begin{tabular}{@{}c@{}}$\sigma_2$\\$[\rm km\,s^{-1}]$\end{tabular}} &
\colhead{\begin{tabular}{@{}c@{}}$v_3$\\$[\rm km\,s^{-1}]$\end{tabular}} &
\colhead{\begin{tabular}{@{}c@{}}$\sigma_3$\\$[\rm km\,s^{-1}]$\end{tabular}}
}
\startdata
8425.97919 & $-$274.439 & 7.769 & 54.416     & 7.274 & \nodata & \nodata \\
8425.98815 & $-$263.807 & 8.267 & 63.591     & 4.943 & \nodata & \nodata \\
8425.99712 & $-$261.090 & 8.494 & 66.758     & 2.963 & \nodata & \nodata \\
8426.00609 & $-$258.914 & 7.182 & 59.953     & 6.349 & \nodata & \nodata \\
8426.01505 & $-$241.155 & 7.440 & 62.502     & 5.422 & \nodata & \nodata \\
8426.02401 & $-$232.721 & 6.895 & 56.430     & 5.855 & \nodata & \nodata \\
8426.03299 & $-$220.081 & 5.865 & 48.353     & 4.358 & \nodata & \nodata \\
8426.04195 & $-$203.561 & 5.318 & 49.455     & 3.821 & \nodata & \nodata \\
8426.05092 & $-$170.225 & 8.995 & 44.338     & 5.919 & \nodata & \nodata \\
8426.05988 & $-$170.730 & 6.836 & 32.662     & 4.471 & \nodata & \nodata \\
8426.06886 & $-$151.222 & 6.394 & 27.526     & 3.949 & \nodata & \nodata \\
8426.07782 & $-$114.652 & 6.131 & 31.288     & 5.089 & \nodata & \nodata \\
8426.08678 & $-$107.763 & 4.900 & 26.967     & 4.584 & \nodata & \nodata \\
8426.16983 & 184.101    & 8.253 & $-$127.280 & 4.488 & \nodata & \nodata \\
8426.17880 & 192.114    & 9.073 & $-$130.098 & 4.290 & \nodata & \nodata \\
8426.18776 & 196.385    & 9.388 & $-$131.226 & 5.175 & \nodata & \nodata \\
8426.19673 & 201.341    & 8.911 & $-$132.098 & 4.574 & \nodata & \nodata \\
8426.20570 & 206.143    & 8.322 & $-$130.763 & 4.546 & \nodata & \nodata \\
8426.21466 & 212.609    & 7.775 & $-$136.459 & 5.748 & \nodata & \nodata \\
8426.22362 & 201.090    & 7.511 & $-$131.453 & 4.602 & \nodata & \nodata \\
8426.23260 & 191.621    & 6.772 & $-$125.036 & 2.828 & \nodata & \nodata \\
8427.21329 & \nodata    & \nodata & $-$16.156  & 4.280 & \nodata & \nodata \\
8427.34956 & $-$238.846 & 5.377 & 49.454     & 6.941 & \nodata & \nodata \\
9211.94190 & 161.632    & 9.205 & $-$117.888 & 7.065 & $-$16.723 & 2.261 \\
9211.95058 & 180.003    & 9.381 & $-$117.723 & 6.095 & $-$15.253 & 1.778 \\
9211.95926 & 187.140    & 5.242 & $-$128.602 & 6.222 & $-$18.535 & 1.730 \\
9211.96794 & 200.892    & 7.606 & $-$124.510 & 6.434 & $-$17.338 & 1.925 \\
9211.97662 & 215.464    & 6.742 & $-$127.063 & 4.906 & $-$17.913 & 1.530 \\
9211.98530 & 216.539    & 8.093 & $-$126.181 & 7.173 & $-$16.254 & 0.899 \\
9211.99398 & 205.106    & 8.957 & $-$133.564 & 5.932 & $-$16.290 & 1.535 \\
9212.00266 & 208.612    & 7.623 & $-$121.182 & 6.710 & $-$16.878 & 2.060 \\
9212.09532 & \nodata    & \nodata & $-$48.040  & 3.957 & $-$16.698 & 0.790 \\
9212.10435 & \nodata    & \nodata & $-$25.068  & 4.163 & $-$15.876 & 0.688 \\
9212.11337 & \nodata    & \nodata & $-$15.005  & 4.282 & $-$16.082 & 0.583 \\
9212.12240 & \nodata    & \nodata & $-$14.475  & 5.082 & $-$16.109 & 0.657 \\
9212.13143 & $-$188.594 & 5.159 & 11.213     & 5.574 & $-$16.444 & 0.874 \\
\enddata
\end{deluxetable}

\begin{deluxetable}{ccccccc}
\tabletypesize{\scriptsize}
\tablewidth{0pt}
\label{tab:MinMax}
\tablecaption{Main photometric parameters of the light curves for BS~Cas.} \label{tab:DifMag}
\tablehead{
\colhead{Date} & 
\colhead{Band} & 
\colhead{\begin{tabular}{@{}c@{}}$\mathrm{Max~I-Min~II}$\\  $[\rm mag]$\end{tabular}} &
\colhead{\begin{tabular}{@{}c@{}}$\mathrm{Max~II-Min~II}$\\ $[\rm mag]$\end{tabular}} & 
\colhead{\begin{tabular}{@{}c@{}}$\mathrm{Min~I-Min~II}$\\  $[\rm mag]$\end{tabular}} & 
\colhead{\begin{tabular}{@{}c@{}}$\mathrm{Max~I-Max~II}$\\  $[\rm mag]$\end{tabular}} &
\colhead{Reference}
}
\startdata
Sep--Dec 2004   & $V$    & $-$0.608 & $-$0.621 & $-$0.079 & 0.013  &OMC       \\[1pt]
\noalign{\vskip 4.5pt}
Jun--Jul 2005        & $V$    & $-$0.607 & $-$0.608 & $-$0.008 & 0.001  &OMC       \\[1pt]
\noalign{\vskip 4.5pt}
Feb 2007             & $V$    & $-$0.612 & $-$0.647 & 0.063  & 0.035  &OMC       \\[1pt]
\noalign{\vskip 4.5pt}
Aug--Dec 2007        & $wide$ & $-$0.550 & $-$0.557 &  0.002  & 0.007 &SuperWASP \\[1pt]
\noalign{\vskip 4.5pt}
Sep--Oct 2007   & $R$    & $-$0.598 & $-$0.603 & $-$0.003 & 0.005  &H10       \\[1pt]
\noalign{\vskip 4.5pt}
Nov 2007--Jan 2008   & $B$    & $-$0.463 & $-$0.387 & 0.159  & $-$0.076 &Y08       \\
                     & $V$    & $-$0.468 & $-$0.431 & 0.103  & $-$0.038 &Y08       \\
                     & $R$    & $-$0.446 & $-$0.402 & 0.102  & $-$0.044 &Y08       \\[1pt]
\noalign{\vskip 4.5pt}
Sep--Oct 2011   & $B$    & $-$0.616 & $-$0.595 & 0.061  & $-$0.021 &CbNUOJ    \\
                     & $V$    & $-$0.589 & $-$0.574 & 0.050  & $-$0.015 &CbNUOJ    \\[1pt]
\noalign{\vskip 4.5pt}
Jul--Aug 2013   & $V$    & $-$0.621 & $-$0.586 & $-$0.045 & $-$0.035 &AAVSO     \\
                     & $R$    & $-$0.597 & $-$0.582 & $-$0.030 & $-$0.015 &AAVSO     \\[1pt]
\noalign{\vskip 4.5pt}
Aug--Sep 2014   & $V$    & $-$0.633 & $-$0.610 & $-$0.016 & $-$0.023 &AAVSO     \\
                     & $R$    & $-$0.615 & $-$0.613 & $-$0.029 & $-$0.001 &AAVSO     \\[1pt]
\noalign{\vskip 4.5pt}
Sep 2014--Feb 2015   & $R$    &  \nodata &  \nodata & $-$0.032 &  \nodata &SOAO      \\[1pt]
\noalign{\vskip 4.5pt}
Sep 2016             & $V$    & $-$0.631 &  \nodata & $-$0.037 &  \nodata &AAVSO     \\[1pt]
\noalign{\vskip 4.5pt}
Nov 2018             & $R$    &  \nodata &  \nodata & $-$0.036 &  \nodata &SOAO      \\[1pt]
\noalign{\vskip 4.5pt}
Nov 2019             & $TESS$ & $-$0.553 & $-$0.555 & $-$0.014 & 0.002    &TESS      \\[1pt]
\noalign{\vskip 4.5pt}
Dec 2020             & $B$    & $-$0.637 & $-$0.653 & $-$0.022 & 0.016    &SOAO      \\
                     & $V$    & $-$0.612 & $-$0.623 & $-$0.015 & 0.011    &SOAO      \\
                     & $R$    & $-$0.597 & $-$0.606 & $-$0.010 & 0.009    &SOAO      \\[1pt]
\noalign{\vskip 4.5pt}
Dec 2020--Feb 2021   & $V$    & $-$0.553 & $-$0.547 &  0.008   & $-$0.007 &OMC       \\[1pt]
\noalign{\vskip 4.5pt}
Oct--Nov 2022   & $TESS$ & $-$0.564 & $-$0.568 & $-$0.004 & $-$0.005 &TESS      \\[1pt]
\noalign{\vskip 4.5pt}
Oct 2024             & $V$    & $-$0.630 & $-$0.613 & $-$0.007 & $-$0.017 &LOAO      \\
                     & $R$    & $-$0.618 & $-$0.602 & $-$0.011 & $-$0.016 &LOAO      \\
\noalign{\vskip 4.5pt}
Oct--Nov 2024   & $TESS$ & $-$0.555 & $-$0.558 & $-$0.006 & $-$0.003 &TESS      \\[1pt]
\enddata
\end{deluxetable}

\begin{deluxetable}{ccc}
\tabletypesize{\scriptsize}
\tablewidth{0pt}
\tablecaption{BS~Cas parameter determined by WD.}
\label{tab:WDsolution}
\tablehead{Parameter & Star~1 & Star~2 }
\startdata
$T_0$ (+2,452,500)        & \multicolumn{2}{c}{$0.35245 \pm0.00528$}\\
$P$ [day]                 & \multicolumn{2}{c}{$0.44046121 \pm0.00000035$}\\
$i$ [$\degr$]          & \multicolumn{2}{c}{$89.860 \pm0.175$}\\
$q$                       & \multicolumn{2}{c}{$2.515 \pm0.003$}\\
$T$ [K]                   &       $6439\pm3$        &       6307\\
$\gamma_{0}$ [$\mathrm{km\,s^{-1}}$]   & \multicolumn{2}{c}{$-35.97\pm0.97$}   \\
$a$ [R$_{\odot}$]         & \multicolumn{2}{c}{$3.064\pm0.027$}   \\
${\Omega }_1={\Omega }_2$ & \multicolumn{2}{c}{$5.769\pm0.003$}  \\
$f$ [\%]                  & \multicolumn{2}{c}{$32.250$}    \\
$\ell_{1,B}/(\ell_{1,B}+\ell_{2,B})$   & \multicolumn{2}{c}{$0.3342\pm0.0008$} \\
$\ell_{1,V}/(\ell_{1,V}+\ell_{2,V})$   & \multicolumn{2}{c}{$0.3272\pm0.0006$} \\
$\ell_{1,R}/(\ell_{1,R}+\ell_{2,R})$  & \multicolumn{2}{c}{$0.3240\pm0.0006$} \\
$\ell_{3,B}$                 & \multicolumn{2}{c}{$0.0528\pm0.0017$}     \\
$\ell_{3,V}$                 & \multicolumn{2}{c}{$0.0368\pm0.0015$}    \\
$\ell_{3,R}$                 & \multicolumn{2}{c}{$0.0277\pm0.0015$}     \\
$r_{\rm{pole}}$                &       $0.2977\pm0.0003$ &  $0.4480\pm0.0002$      \\
$r_{\rm{side}}$                &       $0.3128\pm0.0003$ &  $0.4821\pm0.0003$      \\
$r_{\rm{back}}$                &       $0.3584\pm0.0006$ &  $0.5148\pm0.0004$      \\
$\theta$ [$\degr$]   &  \nodata &    9.9655     \\
$\phi$ [$\degr$]     &  \nodata &    68.0582    \\
$r_{\rm{spot}}$ [$\degr$] &  \nodata &    33.2379    \\
$T_{\rm{spot}}$/$T_{\rm{star}}$   &  \nodata &    0.8258     \\ 
\noalign{\vskip 2pt}
\hline 
\noalign{\vskip 2pt}
$M$ [$\rm{M_{\odot}}$]  &  $0.566\pm0.011$   & $1.423\pm0.031$ \\
$R$ [$\rm{R_{\odot}}$]  &  $0.986\pm0.007$   & $1.473\pm0.010$ \\
$L$ [$\rm{L_{\odot}}$]  &  $1.507\pm0.078$  &  $3.093\pm0.149$ \\
$\mathrm{log}\,g$ [cgs] &  $4.203\pm0.004$  &  $4.255\pm0.004$ 
\enddata
\end{deluxetable}

\begin{deluxetable}{ccccccc}
\tabletypesize{\tiny}
\tablewidth{0pt}
\tablecaption{Spot and $\ell_{3}$ model solutions.}
\label{tab:WDsolution_All}
\tablehead{ & OMC 2004                  & OMC 2005                  & OMC 2007                  & SuperWASP \& H10 2007      & CbNUOJ 2011              & AAVSO 2013              }
\startdata
$T_0\,(+2452500)$          & $0.27821\pm0.00043$       & $0.27957\pm0.00034$       & $0.27915\pm0.00068$       & $0.28518\pm0.00044$       & $0.31123\pm0.00115$      & $0.31128\pm0.00165$     \\
$P\,[\mathrm{day}]$        & \shortstack{$0.44046656$\\$\pm0.00000008$}   & \shortstack{$0.44046646$\\$\pm0.00000008$}   & \shortstack{$0.44046646$\\$\pm0.00000010$}   & \shortstack{$0.44046606$\\$\pm0.00000010$}   & \shortstack{$0.44046414$\\$\pm0.00000015$}   & \shortstack{$0.44046413$\\$\pm0.00000018$} \\
$\ell_{1,B}/(\ell_{1,B}+\ell_{2,B})$  
                           &                           &                           &                           &                           & $0.3342\pm0.0010$        &                         \\
$\ell_{1,V}/(\ell_{1,V}+\ell_{2,V})$  
                           & $0.3272\pm0.0029$         & $0.3272\pm0.0023$         & $0.3272\pm0.0042$         &                           & $0.3272\pm0.0008$        & $0.3272\pm0.0010$       \\
$\ell_{1,R}/(\ell_{1,R}+\ell_{2,R})$  
                           &                           &                           &                           & $0.3232\pm0.0012$         &                          & $0.3232\pm0.0009$       \\
$\ell_{1,wide}/(\ell_{1,wide}+\ell_{2,wide})$
                           &                           &                           &                           & $0.3278\pm0.0014$         &                          &                         \\
$\ell_{3,B}$               &                           &                           &                           &                           & $0.0887\pm0.0019$        &                         \\
$\ell_{3,V}$               & $0.1056\pm0.0067$         & $0.0688\pm0.0056$         & $0.0258\pm0.0100$         &                           & $0.0739\pm0.0017$        & $0.0797\pm0.0023$       \\
$\ell_{3,R}$               &                           &                           &                           & $0.0326\pm0.0028$         &                          & $0.0636\pm0.0021$       \\
$\ell_{3,wide}$            &                           &                           &                           & $0.1009\pm0.0031$         &                          &                         \\
$\theta_1\,[\degr]$       & $9.9655$                  & $9.9655$                  & $14.5491$                 & $11.9364$                 & $10.9756$                & $25.3837$               \\
$\phi_1\,[\degr]$         & $6.2722$                  & $36.8824$                 & $223.952$                 & $28.1288$                 & $213.0887$               & $319.8588$              \\
$r_{\rm{spot,1}}\,[\degr]$     & $37.5001$                 & $26.2438$                 & $39.8274$                 & $18.6423$                 & $56.2983$                & $31.0853$               \\
$T_{\rm{spot,1}}/T_{\rm{star,2}}$    & $0.8258$                  & $0.8258$                  & $0.8258$                  & $0.8562$                  & $0.8314$                 & $0.8241$                \\
$\theta_2\,[\degr]$       &                           &                           &                           &                           & $79.8084$                &                         \\
$\phi_2\,[\degr]$         &                           &                           &                           &                           & $29.2965$                &                         \\
$r_{\rm{spot,2}}\,[\degr]$     &                           &                           &                           &                           & $10.3287$                &                         \\
$T_{\rm{spot,2}}/T_{\rm{star,2}}$    &                           &                           &                           &                           & $0.7934$                 &                         \\ \noalign{\vskip 4.5pt} \hline
               & AAVSO 2014                & AAVSO 2016                & TESS 2019                 & OMC 2020$^{\textstyle *}$       & TESS 2022                 & TESS \& LOAO 2024               \\ \noalign{\vskip 4.5pt} \hline
$T_0\,(+2452500)$          & $0.30895\pm0.00207$       & $0.30276\pm0.00428$       & $0.35513\pm0.00386$       & $0.35245$       & $0.34627\pm0.00145$       & $0.27404\pm0.00040$     \\
$P\,[\mathrm{day}]$ & \shortstack{$0.44046434$\\$\pm0.00000021$} & \shortstack{$0.44046473$\\$\pm0.00000036$} & \shortstack{$0.44046087$\\$\pm0.00000027$} & \shortstack{$0.44046121$\\$ $} & \shortstack{$0.44046687$\\$\pm0.00000006$} & \shortstack{$0.44046261$\\$\pm0.00000016$} \\
$\ell_{1,V}/(\ell_{1,V}+\ell_{2,V})$  
                           & $0.3272\pm0.0009$         & $0.3272\pm0.0008$         &                           & $0.3272\pm0.0004$                  &                           & $0.3273\pm0.0012$       \\
$\ell_{1,R}/(\ell_{1,R}+\ell_{2,R})$  
                           & $0.3232\pm0.0008$         &                           &                           &                           &                           & $0.3240\pm0.0012$       \\
$\ell_{1,TESS}/(\ell_{1,TESS}+\ell_{2,TESS})$
                           &                           &                           & $0.3125\pm0.0002$         &                           & $0.3125\pm0.0001$         & $0.3125\pm0.0016$       \\
$\ell_{3,V}$               & $0.0431\pm0.0022$         & $0.0456\pm0.0018$         &                           & $0.0368         $         &                           & $0.0198\pm0.0028$       \\
$\ell_{3,R}$               & $0.0278\pm0.0020$         &                           &                           &                           &                           & $0.0135\pm0.0029$       \\
$\ell_{3,TESS}$            &                           &                           & $0.0093\pm0.0005$         &                           & $0.0000\pm0.0002$         & $0.0068\pm0.0042$       \\
$\theta_1\,[\degr]$       & $23.9966$                 & $40.6261$                 & $16.2691$                 & $9.9655$                 & $8.8808$                  & $16.034$               \\
$\phi_1\,[\degr]$         & $326.6541$                & $355.8675$                & $54.156$                  & $68.0582$                 & $139.8556$                & $265.6312$              \\
$r_{\rm{spot,1}}\,[\degr]$     & $23.5279$                 & $13.6765$                 & $19.8266$                 & $33.2379$                 & $38.1132$                 & $20.4025$               \\
$T_{\rm{spot,1}}/T_{\rm{star,2}}$     & $0.8241$                  & $0.7859$                  & $0.7766$                  & $0.8258$                  & $0.7766$                 & $0.8441$                 \\
\enddata
\tablenotetext{*}{To match the observed light curve, only $\ell_1$ was adjusted, while all other parameters were fixed to the values of the SOAO2020 light-curve solution listed in Table \ref{tab:WDsolution}, which was obtained at a similar epoch.}
\end{deluxetable}

\begin{deluxetable}{cccccccccc}
\tabletypesize{\tiny}
\tablewidth{0pt}
\tablecaption{Contact binary systems showing eclipse-depth reversal.}
\label{tab:subtype_reversal}
\tablehead{
\colhead{Name} &
\colhead{$P$ [days]} &
\colhead{$dP/dt$ [days~yr$^{-1}$]} &
\colhead{$q$} &
\colhead{$f$ [\%]} &
\colhead{$i$ [$\degr$]} &
\colhead{$T_1$ [K]} &
\colhead{$T_2$ [K]} &
\colhead{subtype} &
\colhead{Reference}
}
\startdata
EK~Com & 0.227 & $3.27 \times 10^{-8}$ & 3.64 & 13.1 & 84.22 & 5354 & 5000 & W & 1 \\
V2790~Ori & 0.288 & $-3.18 \times 10^{-8}$ & 3.10 & 1.1--32.4 & 81.6--88.3 & 5393--5794 & 5314 & W & 2 \\
GN~Boo & 0.302 & $1.74 \times 10^{-7}$ & 3.14 & 14.53--28.29 & 82.26--84.90 & 5310 & 4860--5253 & W & 3 \\
V1191~Cyg & 0.313 & $3.13 \times 10^{-6}$ & 9.36 & 57.9 & 80.76 & 6375 & 6215 & W & 4 \\
FG~Hya & 0.328 & $6.7 \times 10^{-8}$ & 7.657--7.855 & 69--74 & 86.60--87.12 & 5811--5852 & 5900 & A & 5 \\[-1.75ex]
       &       &                           & 8.969 & 85.6 & 82.25 & 6012 & 5900 & W & \\
RZ~Com & 0.339 & $4.12 \times 10^{-8}$ & 1.295 &  8.9  & 78.402 & 4802 & 4900 & A & 6 \\[-1.75ex]
       &       &                       & 2.351 &  20.1 & 81.4   & 5000 & 4900 & W & \\
TY~UMa & 0.355 & $5.18 \times 10^{-7}$ & 2.52 & 13.4 & 84.9 & 6250 & 6229 & W & 7 \\
V410~Aur & 0.366 & $3.42 \times 10^{-8}$$^{\textstyle *}$ & 6.94 &  29 & 81.8--82.5  & 5893--5913 & 5950 & A & 8 \\[-1.75ex]
         &       &                 & 6.94 & 68.6--79.0 & 83.18--83.30 & 5926--6072 & 5760 & W &  \\
AM~Leo & 0.366 & $2.61 \times 10^{-8}$$^{\textstyle *}$ & 2.273 & 22.5 & 89.0 & 6273 & 5942 & W & 9 \\
TYC~4002-2628-1 & 0.367 & an abrupt change & 20.75 & 5--35 & 69.1--69.9 & 6044--6151 & 6032 & W & 10\\
CN~Tri  & 0.367 & no change & 12.92--12.47 & 59.9--82.4 & 71.92--76.21 & 6092--6223 & 5843 & W & 11\\
V752~Cen & 0.370 & $1.61 \times 10^{-7}$ & 3.212 & 25.7--27.6 & 83.05--83.42 & 6220--6360 & 6138 & W & 12 \\
BS~Cas & 0.440 & $-2.30 \times 10^{-7}$ & 2.515 & 32.250 & 89.860 & 6439 & 6307 & W & 13
\enddata

\tablecomments{$^{``*"}$: The values of $dP/dt$ were determined including recent times of minima in this paper}
\tablerefs{(1) \citet{2017NewA...56...14T,2023AJ....165..259L},
(2) \citet{2026ApJ...997..289W}, (3) \citet{2015AJ....149..164W},
(4) \citet{2011AJ....142..124Z,2012NewA...17...46U,2005ApnSS.296..281P,2014NewA...31...14O},
(5) \citet{2000AnAS..144..457Y, 2005MNRAS.356..765Q, 2026MNRAS.tmp..158Y}, (6) \citet{2005PASJ...57..977Q,2008ChJAA...8..465H},
(7) \citet{2002MNRAS.331..707K, 2015AJ....149..120L}, (8) \citet{2022ApJ...927..183L,2017AJ....154...99L},
(9) \citet{2016AstBu..71...64G, 2017IBVS.6227....1G, 2020ARep...64..922G, 1982ApnSS..83..391H}, 
(10) \citet{2022MNRAS.517.1928G}, (11) \citet{2026ApJ...997..270W}, 
(12) \citet{2025NatSR..1527202Y}, (13) This paper.}
\end{deluxetable}

\clearpage

\clearpage

\appendix
		
\section{Time-scaled Light Curves for Collected Photometric Data}\label{sec:app_minima}
\setcounter{table}{0}
\setcounter{figure}{0}
\renewcommand{\thetable}{A\arabic{table}}
\renewcommand{\thefigure}{A\arabic{figure}}
\suppressfloats[t]
The time-scaled TESS light curves for Sectors 18, 58, and 85 are shown in panels (a)--(c) in Figure \ref{fig:LCtess_time}, respectively.

\begin{figure*}[!b]
\epsscale{0.7}
\plotone{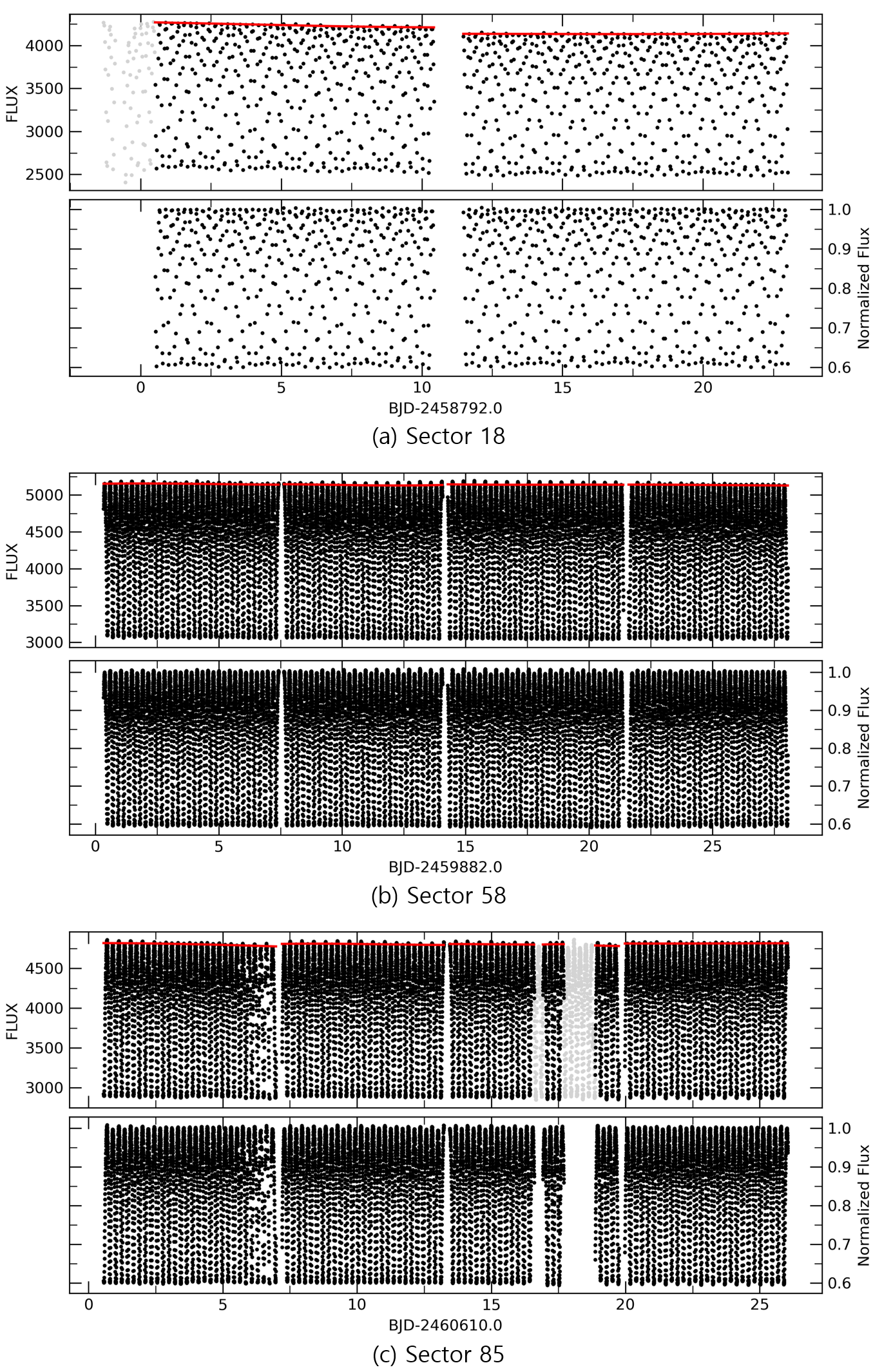}
\caption{
TESS time-series light curves for BS~Cas in Sectors 18, 58, and 85.
Panels (a)--(c) correspond to Sectors 18, 58, and 85, respectively.
In each panel, the upper panel shows the original data extracted from the TPFs, and the lower panel displays the normalized and detrended light curves.
The gray and black dots represent the raw and selected data, respectively.
The red lines denote the spline fits.}
\label{fig:LCtess_time}
\end{figure*}

\end{document}